\documentclass[aps,prd,preprint,eqsecnum,tightenlines,nofootinbib]{revtex4}

\usepackage[dvips]{graphicx}
\usepackage{epsf}
\usepackage{amsfonts}
\usepackage{amsmath, amsthm, amssymb}
\usepackage{slashed}
\usepackage{comment}
\usepackage{subfig}
\usepackage{caption}

\usepackage{color}
\usepackage{soul}
\usepackage[normalem]{ulem}

\def\fig#1{Fig.~{\ref{#1}}}
\def\Fig#1{Fig.~{\ref{#1}}}
\def\figs#1#2{Figs.~{\ref{#1}} and {\ref{#2}}}

\def\eqn#1{Eq.~({\ref{#1}})}
 
\def\sect#1{Section~{\ref{#1}}}
\def\tab#1{Table~{\ref{#1}}}

\def\app#1{Appendix~{\ref{#1}}}

\begin{document}

\title{
\Large Color-Kinematics Duality in One-Loop Four-Gluon Amplitudes with Matter
}
 
\author{Josh~Nohle}

\affiliation{
Department of Physics and Astronomy, University of California 
at Los Angeles\\ 
 Los Angeles, CA 90095-1547, USA \\ 
$\null$ \\
}

\vskip .5 cm

\begin{abstract}
Four-point one-loop nonsupersymmetric pure Yang-Mills amplitudes
with the duality between color and kinematics manifest have been
constructed in previous work. Here, we extend the discussion to
fermions and scalars circulating in the loop with all external
gluons.  This gives another nontrivial loop-level example showing that
the duality between color and kinematics holds in nonsupersymmetric gauge 
theory.  The construction is valid in any spacetime dimension and
written in terms of formal polarization vectors. We also convert these
expressions into a four-dimensional form with explicit external
helicity states. Using this, we compare our results to one-loop duality-satisfying 
amplitudes that are already present in literature.
\end{abstract}


\maketitle

%%%%%%%%%%%%%%%%%%%%%%%%%%%%%%%%%%%%%%%%%%%%
\section{Introduction}
Recently, a duality between color and kinematics in gauge
theories---also known as the Bern-Carrasco-Johansson (BCJ) duality---has
been uncovered~\cite{BCJ,BCJLoop}. This duality has proven very useful
for constructing multi-loop amplitudes in (super)gravity and for
studying their ultraviolet
properties~\cite{BCJLoop, OneTwoLoopN4,
  SchnitzerBCJ, ck4l, TwoloopHalfMax, ThreeloopHalfMax, nonSUSYBCJ,
  Matter,N4FourLoops}. Specifically, loop integrands in theories of (super)gravity
are obtained from gauge-theory integrands simply by replacing color
factors with duality-satisfying kinematic numerator factors, called
BCJ numerators. This is known as the ``double-copy'' construction,
shown to be valid in Ref.~\cite{GravYM2}.  

At tree level, explicit forms of $n$-point, $D$-dimensional amplitudes
satisfying the duality have been found~\cite{TreeAllN}. At loop
level, the duality remains a conjecture, but there is already
nontrivial evidence in its favor, especially for supersymmetric
theories~\cite{BCJLoop, ck4l, N4Five, Oxidation, OConnellRational,
  BoelsFormFactor, OneLoopN1Susy, nonSUSYBCJ}. Beyond these explicit
duality-satisfying constructions, the duality implies nontrivial
relations amongst gauge-theory color-ordered partial tree
amplitudes~\cite{BCJ,BCJProofs}.  There is also an partial
understanding of the duality at the level of the
Lagrangian~\cite{GravYM2,OConnell,Weinzierl}.  The BCJ duality also points to new
hidden symmetries. In particular, in the self-dual case an underlying
infinite-dimensional Lie algebra corresponding to area preserving
diffeomorphisms has been shown to be responsible for the
duality~\cite{OConnell,OConnellAlgebras}.  Even after carrying out
loop integrations, the duality points to strong links between gravity
and gauge-theory amplitudes~\cite{OneTwoLoopN4, SchnitzerBCJ,
  ck4l,TwoloopHalfMax,Matter}.

The duality was noticed long ago for four-point tree-level Feynman
diagrams as a possible way to explain certain radiation
zeros~\cite{Halzen}. For higher points or at loop level, the duality is
rather nontrivial and no longer holds for ordinary Feynman diagrams,
but requires nontrivial rearrangements to display it. 

At loop level, the duality between color and kinematics has been
extensively studied for supersymmetric cases but less so for the
nonsupersymmetric case.  In particular, four-point one-loop amplitudes
in nonsupersymmetric ($\mathcal{N}=0$) Yang-Mills (YM) theory that
satisfy the BCJ duality have been constructed in
Ref.~\cite{nonSUSYBCJ}. They are valid in arbitrary dimensions and
written in terms of formal polarization vectors---i.e., the external
states are not in a helicity basis. The $\mathcal{N}=0$ YM theory of
Ref.~\cite{nonSUSYBCJ} contains only gluons. Here, we extend that work
by constructing duality-satisfying amplitudes that are valid for any
adjoint matter content circulating in the loop, still with external
gluons. The construction closely follows that of
Ref.~\cite{nonSUSYBCJ}. We begin by building an ansatz for the
amplitude in $D$ dimensions. We use formal polarization vectors
instead of dimension-specific helicity states. The ansatz is then
constrained to satisfy the BCJ duality. Furthermore, we demand that the
kinematic numerator factors of the ansatz obey the same relabeling
symmetries as their corresponding diagrams. Using a $D$-dimensional
variant~\cite{BernMorgan, DDimUnitarity} of the unitarity
method~\cite{UnitarityMethod}, we enforce that the ansatz obey the
same unitarity cuts as the amplitude under consideration. These
$D$-dimensional unitarity cuts completely determine the integrated
amplitude.

Because we have both gluon- and fermion-loop contributions, we can
compare our results to previously obtained one-loop duality-satisfying
amplitudes with four-dimensional external helicity states.  Namely, we
look at the $\mathcal{N}=4$ super-Yang-Mills (sYM) amplitude of
Ref.~\cite{N4}, the maximally-helicity-violating (MHV) $\mathcal{N}=1$
(chiral) sYM amplitude of Ref.~\cite{JJRadu}, and the
all-plus-helicity $\mathcal{N}=0$ YM amplitude of
Ref.~\cite{BernMorgan}. We compare our results to the earlier work by
restricting the external states to four dimensions and putting the formal
polarization vectors into a helicity basis.  While going to a helicity
basis considerably simplifies our MHV $\mathcal{N}=1$ (chiral) BCJ
numerators, the all-plus-helicity $\mathcal{N}=0$ numerators
are not particularly simplified, because they contain complicated terms
that integrate to zero.

We organize the paper as follows. 
In \sect{sec:BCJReview}, we briefly review (1) the anatomy of gauge-theory amplitudes, (2) the conjectured duality between color and kinematics, and (3) how amplitudes in certain theories of (super)gravity are constructed as a double-copy of (super-)Yang-Mills theories. 
In \sect{sec:formal}, we present the BCJ numerators with adjoint matter content circulating in the loop. These individual contributions are then combined to construct BCJ numerators for the theories of $\mathcal{N}=4$ sYM, $\mathcal{N}=1$ (chiral) sYM, and $\mathcal{N}=0$ YM. We show the simplification in combining our formal-polarization BCJ numerators into the four-dimensional supersymmetric theories of $\mathcal{N}=4$ sYM and $\mathcal{N}=1$ (chiral) sYM in \app{sec:num4}.
In \sect{sec:MomBasis}, we review one technique for putting formal polarization vectors into a helicity basis, with a slight digression found in \app{sec:phase}.
In \sect{sec:NumComp}, we  compare our BCJ numerators with the existing literature.
Finally, in \sect{sec:Conclusions}, we present our conclusions.
%%%%%%%%%%%%%%%%%%%%%%%%%%%%%%%%%%%%%%%

%%%%%%%%%%%%%%%%%%%%%%%%%%%%%%%%%%%%%%%
\section{Review}
\label{sec:BCJReview}

An $m$-point $L$-loop gauge-theory amplitude in $D$ dimensions, with
all particles in the adjoint representation, may be written as
\begin{align}
\mathcal{A}^{L\hbox{-}\mathrm{loop}}_{m}=i^{L} g^{m-2+2 L}\sum_{\mathcal{S}_{m}}\sum_{j}\int\prod_{l=1}^{L}\frac{d^{D}p_{l}}{(2\pi)^{D}}\frac{1}{S_{j}}\frac{c_{j} n_{j}}{\prod_{\alpha_{j}}p^{2}_{\alpha_{j}}}\,,
\label{CubicRepresentation}
\end{align}
where $g$ is the gauge coupling constant. The first sum runs over the $m!$ permutations of the external legs, denoted by $\mathcal{S}_{m}$. The $j$-sum is over the set of distinct, nonisomorphic, $m$-point $L$-loop graphs with only cubic vertices. Since a diagram can be identified by its propagators, any diagram with quartic or higher vertices can be converted to a diagram with only cubic vertices by multiplying and dividing by the appropriate propagators, i.e., factors that go as $p_{\alpha}^{2}/p_{\alpha}^{2}$.  The loop integral is over $L$ independent $D$-dimensional loop momenta, $p_{l}$. Associated with graph $j$ are the following:
\begin{itemize}

\item $S_{j}$: The symmetry factor that removes any overcounting from permutations of external legs and also from any internal automorphism symmetries of the graph.

\item $1/\prod_{\alpha_{j}} p^{2}_{\alpha_{j}}$: The propagators affiliated with the graph.

\item $n_{j}$: The numerator that contains the nontrivial kinematic information, dependent on momenta, polarizations, and spinors. (If a superspace form is used in supersymmetric cases, it will depend also on Grassmann parameters.)

\item $c_{j}$: The color factor obtained by dressing every vertex of the graph with the group-theory structure constant, $\tilde{f}^{abc}=i\sqrt{2}f^{abc}=\mathrm{Tr}([T^{a},T^{b}] T^{c})$, where the hermitian generators of the gauge group are normalized via $\mathrm{Tr}(T^{a}T^{b})=\delta^{a b}.$

\end{itemize}
%%%%%%%%%%%%%% FIGURE %%%%%%%%%%%%%
\begin{figure}
\includegraphics[scale=.45]{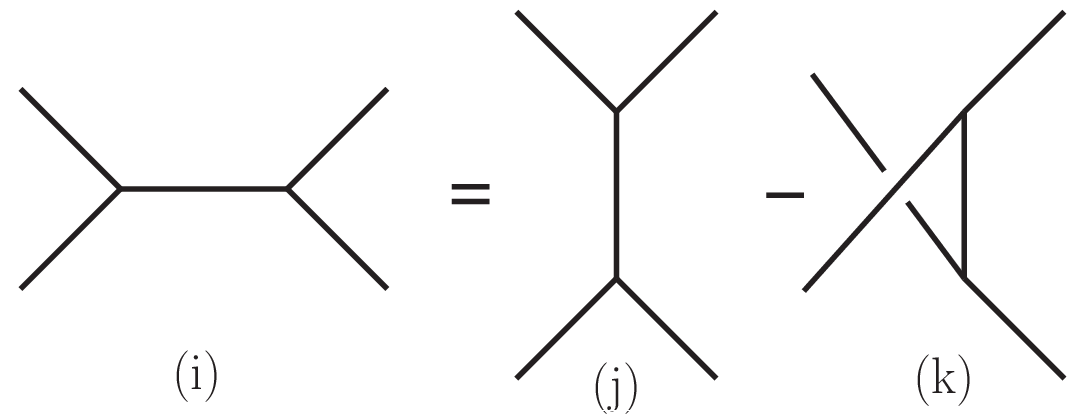}
\caption{The basic Jacobi relation for either color or numerator
  factors.  These three diagrams can be embedded in a larger (loop) diagram. }
\label{fig:BCJ}
\end{figure}
%%%%%%%%%%%%%%%%%%%%%%%%%%%
We briefly mention that the four-point one-loop amplitudes of \eqn{CubicRepresentation}---which we will be concerned with in this paper---can be written as~\cite{DDM}:
\begin{align}
\label{eq:CO}
\mathcal{A}_{4}^{(1)}(1,2,3,4)=g^{4}\left[c^{(1)}_{1234}A_{4}^{(1)}(1,2,3,4)+c^{(1)}_{1423}A_{4}^{(1)}(1,4,2,3)+c^{(1)}_{1342}A_{4}^{(1)}(1,3,4,2)\right].
\end{align}
The $A_{4}^{(1)}$'s are the one-loop color-ordered amplitudes~\cite{ColorDecomp}, which are independently gauge invariant. The color factors, $c^{(1)}_{1i_{2}i_{3}i_{4}}$, are given by dressing the vertices of the one-loop box diagram that has the external-leg ordering $(1,i_{2},i_{3},i_{4})$ with the structure constants $\tilde{f}^{abc}$.

The numerators appearing in \eqn{CubicRepresentation} are by no means unique because of the freedom to move terms between different diagrams, also known as generalized gauge invariance~\cite{BCJ,TyeBCJ,BCJLoop,GravYM2}. The BCJ conjecture is that, to all loop orders, this freedom can be utilized to find representations of the amplitude where the kinematic numerators obey the same algebraic relations that the color factors obey~\cite{BCJ,BCJLoop}. In ordinary gauge theories, this is simply the Jacobi identity,
\begin{equation}
c_{i}=c_{j}-c_{k}\ \Rightarrow\ n_{i}=n_{j}-n_{k} \,,
\label{BCJDuality}
\end{equation}
where $i$, $j$, and $k$ label three diagrams whose color factors obey the Jacobi identity.  The basic Jacobi relation is displayed in \fig{fig:BCJ}.  The generalization of the identity to $m$-point $L$-loop amplitudes is seen diagrammatically by embedding \fig{fig:BCJ} in larger diagrams, where the other parts of the three diagrams remain unaltered. Furthermore, if the color factor of a diagram is antisymmetric under a swap of legs, we require that the numerator obey the same antisymmetry,
\begin{equation}
c_{i} \rightarrow - c_{i}\ \Rightarrow\ n_{i} \rightarrow -n_{i} \,.
\label{BCJFlipSymmetry}
\end{equation}
We note that the numerator relations are nontrivial functional relations because they depend on momenta, polarizations, and spinors, as discussed in some detail in Refs.~\cite{ck4l, JJHenrikReview,JJ2}.

Although we do not focus on gravity amplitudes in this paper, we
briefly mention their construction via the color-kinematics
duality. Once a gauge-theory integrand is constructed with the 
color-kinematics duality manifest, gravity loop integrands are
obtained almost trivially~\cite{BCJ,BCJLoop}.  One simply replaces the
color factor in the gauge-theory integrand with a kinematic numerator
from a second gauge theory,
\begin{equation}
c_{i}\ \rightarrow\ \tilde{n}_{i}\,.
\label{ColorSubstitution}
\end{equation}
This leads to an expression of the gravity amplitude as a double-copy of Yang-Mills theory:
\begin{align}
\mathcal{M}^{L\hbox{-}\mathrm{loop}}_{m} = i^{L+1} \left(\frac{\kappa}{2}\right)^{m-2+2 L}
\sum_{\mathcal{S}_{m}}\sum_{j}\int\prod_{l=1}^{L}\frac{d^{D}p_{l}}{(2\pi)^{D}}
\frac{1}{S_{j}}\frac{\tilde{n}_{j}n_{j}}{\prod_{\alpha_{j}}p^{2}_{\alpha_{j}}} \,,
\label{DoubleCopy}
\end{align}
where $\tilde{n}_j$ and $n_j$ are gauge-theory numerator factors.  Only one of the two sets of numerators needs to satisfy the duality of \eqn{BCJDuality}
~\cite{BCJLoop,GravYM2}.  The double-copy
formalism has been studied at loop level in some detail in a variety
of cases~\cite{BCJLoop,OneTwoLoopN4,N4Five,ck4l, ThreeloopHalfMax,
  TwoloopHalfMax, White, Oxidation}.
%%%%%%%%%%%%%%%%%%%%%%%%%%%%%%%%%%%%%%

%%%%%%%%%%%%%%%%%%%%%%%%%%%%%%%%%%%%%%
\section{Formal-Polarization BCJ Numerators}
\label{sec:formal}
 %%%%%%%%%%%%%% FIGURE %%%%%%%%%%%%%
%%%Match last paper
 \begin{figure}
    \subfloat[]{%
      \includegraphics[scale=.7]{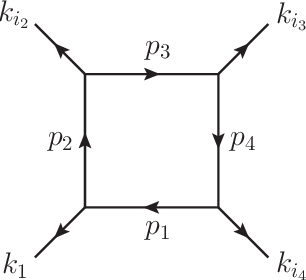}
    }
    \hspace{.75cm}
    \subfloat[]{%
      \includegraphics[scale=.7]{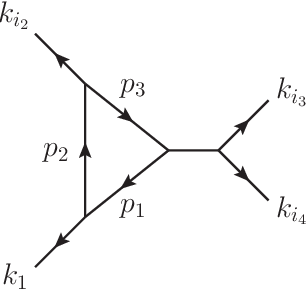}
    }
    \hspace{.75cm}
    \subfloat[]{%
      \includegraphics[scale=.7]{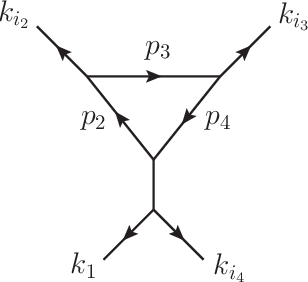}
    }
    \\[.5cm]
    \subfloat[]{%
      \includegraphics[scale=.7]{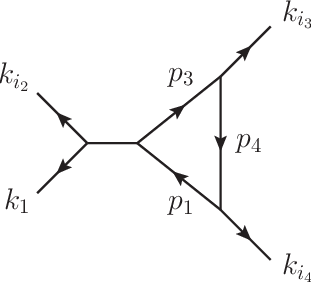}
    }
    \hspace{.75cm}
    \subfloat[]{%
      \includegraphics[scale=.7]{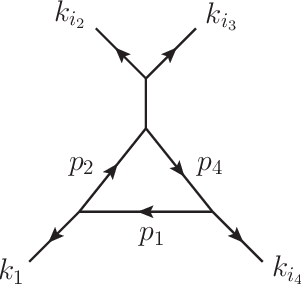}
    }
    \hspace{.75cm}
    \subfloat[]{%
       \raisebox{.65cm}{\includegraphics[scale=.7]{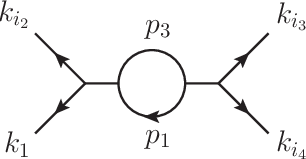}}
    }
    \hspace{.75cm}
    \subfloat[]{%
      \includegraphics[scale=.7]{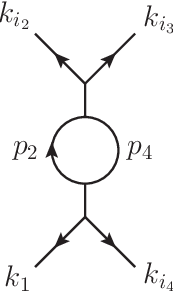}
    }
     \caption{The labeling convention that we employ both for numerators with formal polarization vectors and for color-ordered amplitudes. The external legs have the ordering  $(1, i_{2}, i_{3}, i_{4})$, with outgoing momenta $k_{1}$, $k_{i_{2}}$, $k_{i_{3}}$, $k_{i_{4}}$. The loop momentum is denoted by $p_{1}\equiv p$, while $p_{2}$, $p_{3}$, and $p_{4}$ are given by momentum conservation.}
    \label{fig:ampLabel}
  \end{figure}
\begin{figure}
\includegraphics[scale=.75]{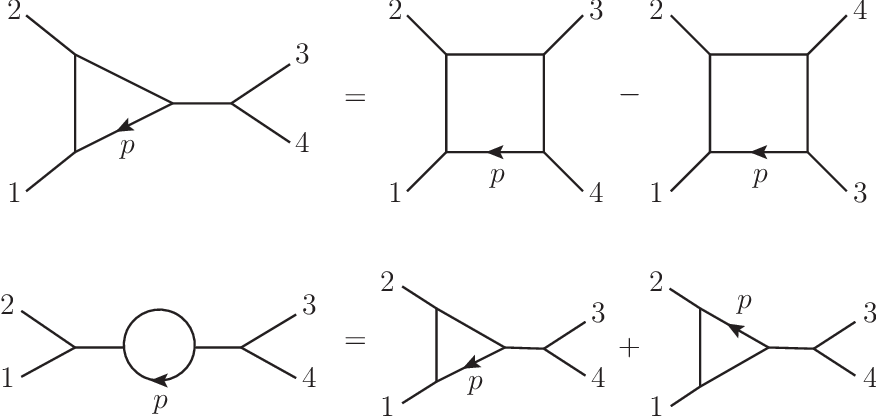}
\caption{The Jacobi relations determining either color or kinematic 
numerators of the four-point diagrams containing either a triangle  
or internal bubble. }
\label{fig:TriBub}
\end{figure}

 %%%%%%%%%%%%%%%%%%%%%%%%%%%%%%%%
In this section, we find the BCJ numerators for adjoint fermions and adjoint scalars circulating in the four-point one-loop box diagram---\fig{fig:ampLabel}(a)---with external gluons. For completeness, we also provide the expression for a gluon in the loop. The box numerators that we give are for the external-leg ordering $(1,2,3,4)$ and with the loop momentum labeling convention $p_{1}\equiv p$, where $p_{1}$ is shown in \fig{fig:ampLabel}(a). The other BCJ numerators, such as those displayed in Figs.~\ref{fig:ampLabel}(b-g), are found by  solving the numerator Jacobi relations of \eqn{BCJDuality}. \figs{fig:TriBub}{fig:SnailTadpole} show the Jacobi relations diagrammatically. We note that the right-hand sides of \figs{fig:TriBub}{fig:SnailTadpole} can be written solely in terms of boxes. In these functional numerator relations, we encounter box numerators with different external-leg orderings and loop-momentum labels. However, we demand that these numerators are simply relabelings of the box numerator that we give. (In this procedure, the polarization vectors must of course be relabeled in addition to the external momenta and the loop momentum.) We also demand that the box numerator is unchanged under the three rotation relabelings and four reflection relabelings---the automorphisms of the box diagram. The other numerators have analogous relabeling properties, which follow from the color-kinematics duality.
%%%%%%%%%%%%%% FIGURE %%%%%%%%%%%%%
%%%%%%%%%%%%%%%%%%%%%%%%%%%
%%%%%%%%%%%%%%% FIGURE %%%%%%%%%%%%
\begin{figure}
\includegraphics[scale=.75]{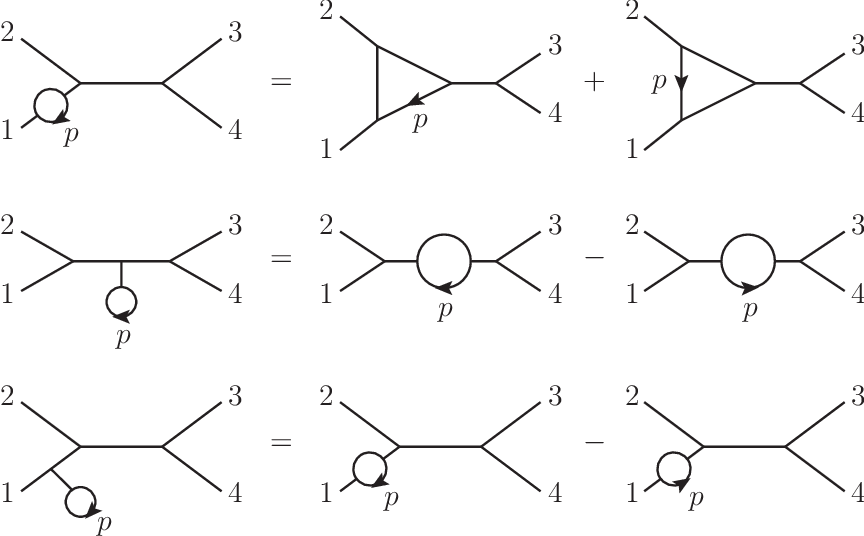}
\caption{The color or kinematic Jacobi relations involving a bubble
  on an external leg or a tadpole. These diagrams have vanishing
  contribution to the integrated amplitude.}
\label{fig:SnailTadpole}
\end{figure}
%%%%%%%%%%%%%%%%%%%%%%%%%%%

\begin{figure}
    \subfloat[]{%
      \raisebox{.5cm}{\includegraphics[scale=.7]{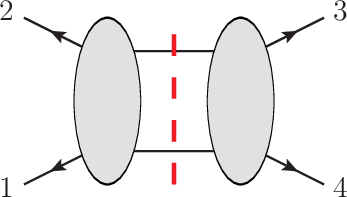}}
    }
    \hspace{1.5cm}
    \subfloat[]{%
       \includegraphics[scale=.62]{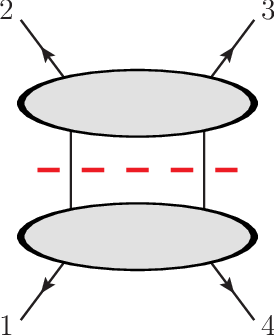}
      }
\caption{The two two-particle unitarity cuts in which the exposed internal propagators are put on shell. The one-loop contributions to cut (a) come from Figs.~\ref{fig:ampLabel}(a,b,d,f), and the one-loop contributions to cut (b) come from Figs.~\ref{fig:ampLabel}(a,c,e,g). Diagrams with a bubble on an external leg and diagrams that contain a tadpole do not contribute to either cut.}
\label{fig:cuts}
\end{figure}

The construction of these BCJ numerators closely follows that of Ref.~\cite{nonSUSYBCJ}. As we will discuss, we generalize the constraints of Ref.~\cite{nonSUSYBCJ} to accommodate matter in the loop, and we also make additional constraints on internal bubble numerators (Figs.~\ref{fig:ampLabel}(f,g)) with supersymmetry in mind. First, we build an ansatz for the box numerator with external-leg ordering $(1,2,3,4)$. The ansatz is a sum of all (\emph{468}) possible terms, each with an undetermined coefficient. Next, we impose the color-kinematics duality and the relabeling properties mentioned above.  This allows us to generate the other numerators needed to construct the color-ordered amplitudes of \eqn{eq:CO}. Then, we enforce that these amplitudes obey the appropriate two-particle $D$-dimensional unitarity cuts of \fig{fig:cuts}. \fig{fig:ampLabel} displays the seven diagrams that contribute to at least one of the two two-particle unitarity cuts of the color-ordered amplitude $A_{4}^{(1)}(1,i_{2},i_{3},i_{4})$. Because of our relabeling properties, we need only to consider one of the color-ordered amplitudes, say $A_{4}^{(1)}(1,2,3,4)$. Imposing the duality, relabeling, and cut conditions fixes \emph{447} of the 468  %%%%%%%%%%%%%% FIGURE %%%%%%%%%%%%%
%%%%%%%%%%%%%%%%%%%%%%%%%%%%%%%%

To clean up the expression, we the fix \emph{12} of the remaining 21
coefficients by demanding that all tadpole numerators vanish prior to
integration. That is, the left-hand side of the bottom two equations
of \Fig{fig:SnailTadpole} are set to zero. (In fact, solving just the
bottom equation in the figure is sufficient.) Because the tadpole
integrals are scale-free in dimensional regularization, they vanish
regardless of the coefficient choice (see Ref.~\cite{Smirnov}). An
important benefit of imposing that the tadpole numerators vanish
\emph{prior} to integration is that the maximum power of loop momentum
in each BCJ numerator is $p^{V}$, where $V$ is the number of vertices
in the loop. When supersymmetry is present, the maximum power is
reduced to no more than $p^{V-2}$, with $V-2\geq0$. (At one loop, this
well-known improved power counting can be seen by using the
second-order formalism for the fermion loop~\cite{Morgan} and the
background-field gauge for the gluon loop.)

We now fix \emph{four} additional coefficients so that the integrals arising
from the diagram with a bubble on external-leg 1---the first diagram of
\fig{fig:SnailTadpole}---are well-defined. (Our relabeling properties
ensure that the integrals from bubbles on different external legs are
also be well-defined.) In general, in the on-shell limit the 
intermediate propagator, $1/k_{1}^{2}\sim1/0$, can cause the the
integrals to be ill-defined. Feynman diagrams avoid 
this because each term in the bubble-on-external-leg-$1$ Feynman
kinematic numerator contains at least one of the following scalar
products
\begin{align}
\label{eq:GoodTerms}
p\cdot \varepsilon_{1}, \hspace{1cm} p\cdot k_{1}, \hspace{1cm} p^{2}.
\end{align}
This constraint along with the associated current conservation
of the vacuum polarization ensures that the power of $k_{1}^{2}$ after
integration is no lower than $(k_{1}^{2})^{(D-4)/2}$. (In these
integrals, we use the prescription where $k_{1}^{2}$ is not put on
shell until the end of the calculation.) With no powers of $k_{1}^{2}$
in the denominator for $D\geq 4$, the expression is
well-defined. Thus, we demand that each term in the
bubble-on-external-leg-$1$ BCJ kinematic numerator contains at least
one of these scalar products, following the structure found 
with ordinary Feynman-gauge Feynman diagrams. We now expound on this subtle
restriction (see also Ref.~\cite{nonSUSYBCJ}).

\Fig{fig:SnailTadpole} shows that there are no terms with an odd power of loop momentum in the bubble-on-external-leg-1 numerator due to the vanishing-tadpole condition. We choose coefficients so as to eliminate terms with no loop momentum. Now, only terms quadratic in the loop momentum remain, and they in fact contain at least one of the scalar products of \eqn{eq:GoodTerms}. By Lorentz invariance, we have
\begin{align}
\label{eq:LI}
\int\frac{d^{D}p}{(2\pi)^{D}}\frac{p^{\mu}\:p^{\nu}}{k_{1}^{2}\:p^{2}\:(p-k_{1})^{2}} = \frac{1}{k_{1}^{2}}\left(g^{\mu\nu}k_{1}^{2}\:A + k_{1}^{\mu}k_{1}^{\nu}\:B\right),
\end{align}
where $A$ and $B$ are scalar integrals. If one of the loop momentum
vectors of \eqn{eq:LI} is contracted with $\varepsilon_{1}$, then
$k_{1}\cdot \varepsilon_{1}$ appears in the prefactor of integral
$B$. This vanishes immediately, so these terms cause no problem. Aside
from $k_{1}\cdot \varepsilon_{1}$, simple power counting (and noting
that $k_{1}^{2}$ is the only scale in the integral) reveals that all
other terms after integration are at least proportional to
$(k_{1}^{2})^{(D-4)/2}$, as mentioned above. So, the integrals clearly
vanish for $D>4$. In $D=4$, the integrals are now well-defined and
vanish in dimensional regularization. We do note that the integral
\begin{align}
\label{eq:bubInt}
\int\frac{d^{D}p}{(2\pi)^{D}}\frac{k_{1}^{2}}{k_{1}^{2}\:p^{2}\:(p-k_{1})^{2}}=\int\frac{d^{D}p}{(2\pi)^{D}}\frac{1}{p^{2}\:(p-k_{1})^{2}}
\end{align}
vanishes through a cancellation of UV and collinear singularities (see
Ref.~\cite{Smirnov}). Thus, these integrals need to be included when
calculating UV divergences in four-dimensional Yang-Mills
theory.  It is interesting to note that in
the corresponding gravity numerator of \eqn{DoubleCopy}, we ensure
that there is an extra scalar product from \eqn{eq:GoodTerms} in each
term from multiplying two Yang-Mills numerators. Thus, in the gravity 
case, the integrals
either vanish due to $k_{1}\cdot \varepsilon_{1}$ or go as
$(k_{1}^{2})^{(D-2)/2}$, which also vanishes with no contribution to the 
UV divergence.

These tadpole and bubble-on-external-leg constraints generalize those
of Ref.~\cite{nonSUSYBCJ} to deal with matter content in the
loop. Unlike Ref.~\cite{nonSUSYBCJ} where the remaining \emph{five} coefficients
are simply set to zero, here we add the additional simplifying
constraint that the terms without loop momentum vanish in the bubble
numerator of Fig.~\ref{fig:ampLabel}(f). Internal bubble and
bubble-on-external-leg numerators now have no $\mathcal{O}(p^{0})$
terms, so they vanish in supersymmetric theories due to the reduced
maximum power of the loop momentum. The internal bubble and
bubble-on-external-leg conditions are also necessary for the
further-improved loop-momentum power counting in
maximally-supersymmetric Yang-Mills theory (e.g., $\mathcal{N}=4$ in
$D=4$ or $\mathcal{N}=1$ in $D=10$). Namely, the maximum power is
$p^{V-4}$, where $V-4\geq0$. This means that triangle and bubble
numerators vanish identically and that box numerators have no powers
of loop momentum. Henceforth, we fix all 468 coefficients using the
above constraints.

Because the Jacobi relations are linear, the linear combinations of BCJ box numerators also obey color-kinematics duality. Thus, we decompose the BCJ box numerator as follows:
\begin{align}
n_{1234;p}=N_{g}\:n^{\text{(gluon)}}_{1234;p}+N_{\!f}\:n^{\text{(fermion)}}_{1234;p}+N_{s}\:n^{\text{(scalar)}}_{1234;p},
\label{eq:numDecomp}
\end{align}
where $1234$ refers to the external-leg ordering and $p$ is the loop momentum. $n^{\text{(gluon)}}_{1234;p}$, $n^{\text{(fermion)}}_{1234;p}$, and $n^{\text{(scalar)}}_{1234;p}$ are the BCJ box numerators corresponding to individual field contributions in the loop, which we provide below. The prefactors $N_{g}$, $N_{\!f}$, and $N_{s}$ are the number of gluons, fermions, and real scalars, respectively, circulating in the loop. For instance, the allowed field content for supersymmetric theories in four dimensions is given in \tab{tab:susy4}. 
%%%%%%%%%%%%%%%Table%%%%%%%%%%%%%%%%%%%%
\begin{table}
  \begin{tabular}{ l || c | c | c }
  &$\phantom{--}N_{g}\phantom{--}$ & $\phantom{--}N_{\!f}\phantom{--}$ &\phantom{ }$ N_{s}\text{(real)}$\\
    \hline \hline
 
 $\mathcal{N}=4$ &
    	 $1$&
	 $4$&
	 $6$\\ \hline
	 
$\mathcal{N}=2$ &
    	 $1$&
	 $2$&
	 $2$\\ \hline
	 
$\mathcal{N}=1\text{ \small(vector)}$ &
    	 $1$&
	 $1$&
	 $0$\\ \hline
	 
$\mathcal{N}=1\text{ \small(chiral)}$ &
    	 $0$&
	 $1$&
	 $2$\\ \hline
    \end{tabular}
\caption{Four-dimensional supersymmetric field content.}
\label{tab:susy4}
\end{table}
%%%%%%%%%%%%%%%%%%%%%%%%%%%%%%%%%%%%%%%%%%

Using the shorthand notation,
\begin{equation}
\begin{gathered}
p_1=p, \hspace{1.cm}
p_2=p-k_1, \hspace{1.cm}
p_3=p-k_1-k_2, \hspace{1.cm}
p_4=p+k_4, \\[.1cm]
\mathcal{E}_{ij}=\varepsilon_i\cdot\varepsilon_j, \hspace{2cm} 
\mathcal{P}_{ij}=p_i\cdot\varepsilon_j, \hspace{2cm} 
\mathcal{K}_{ij} =k_i\cdot\varepsilon_j,
\end{gathered} 
\end{equation}
and using the labeling convention of \fig{fig:ampLabel}(a), the contribution from a real scalar field circulating in the loop is as follows:
%%%%%%Scalar Loop%%%%%%
\begin{align}
 \lefteqn{ \hskip 0 cm 
n^{\text{(scalar)}}_{1234;p}\: =  -i \Bigl[
-\tfrac{1}{24}\:\mathcal{E}_{12}\:\mathcal{E}_{34}\:p^{2}_1\:p^{2}_3
+\tfrac{1}{24}\:\mathcal{E}_{13}\:\mathcal{E}_{24}\:p^{2}_1\:p^{2}_3
+\tfrac{1}{8}\:\mathcal{E}_{14}\:\mathcal{E}_{23}\:p^{2}_1\:p^{2}_3
+\tfrac{1}{6}\:\mathcal{E}_{12}\:\mathcal{P}_{33}\:\mathcal{P}_{44}\:p^{2}_1}
\vspace{.2cm} \nonumber \\ 
& \null
-\tfrac{1}{6}\:\mathcal{E}_{13}\:\mathcal{P}_{22}\:\mathcal{P}_{44}\:p^{2}_1
-\mathcal{E}_{14}\:\mathcal{P}_{22}\:\mathcal{P}_{33}\:p^{2}_1
-\tfrac{1}{6}\:\mathcal{E}_{24}\:\mathcal{P}_{11}\:\mathcal{P}_{33}\:p^{2}_1
+\tfrac{1}{6}\:\mathcal{E}_{34}\:\mathcal{P}_{11}\:\mathcal{P}_{22}\:p^{2}_1
\vspace{.2cm} \nonumber \\ 
& \null
+\mathcal{P}_{11}\:\mathcal{P}_{22}\:\mathcal{P}_{33}\:\mathcal{P}_{44}
-\tfrac{1}{12}\:\mathcal{E}_{13}\:\mathcal{K}_{42}\:\mathcal{P}_{44}\:p^{2}_1
-\tfrac{1}{12}\:\mathcal{E}_{23}\:\mathcal{K}_{41}\:\mathcal{P}_{44}\:p^{2}_1
+\tfrac{1}{12}\:\mathcal{E}_{12}\:\mathcal{K}_{23}\:\mathcal{P}_{44}\:p^{2}_1
\vspace{.2cm} \nonumber \\ 
& \null
-\tfrac{1}{12}\:\mathcal{E}_{13}\:\mathcal{K}_{12}\:\mathcal{P}_{44}\:p^{2}_1
+\tfrac{1}{12}\:\mathcal{E}_{14}\:\mathcal{K}_{42}\:\mathcal{P}_{33}\:p^{2}_1
-\tfrac{1}{12}\:\mathcal{E}_{24}\:\mathcal{K}_{41}\:\mathcal{P}_{33}\:p^{2}_1
+\tfrac{1}{4}\:\mathcal{E}_{12}\:\mathcal{K}_{34}\:\mathcal{P}_{33}\:p^{2}_1
\vspace{.2cm} \nonumber \\ 
& \null
-\tfrac{1}{6}\:\mathcal{E}_{24}\:\mathcal{K}_{31}\:\mathcal{P}_{33}\:p^{2}_1
+\tfrac{1}{12}\:\mathcal{E}_{12}\:\mathcal{K}_{24}\:\mathcal{P}_{33}\:p^{2}_1
-\tfrac{1}{12}\:\mathcal{E}_{14}\:\mathcal{K}_{12}\:\mathcal{P}_{33}\:p^{2}_1
+\tfrac{1}{4}\:\mathcal{E}_{34}\:\mathcal{K}_{41}\:\mathcal{P}_{22}\:p^{2}_1
\vspace{.2cm} \nonumber \\ 
& \null
-\tfrac{1}{12}\:\mathcal{E}_{13}\:\mathcal{K}_{34}\:\mathcal{P}_{22}\:p^{2}_1
+\tfrac{1}{6}\:\mathcal{E}_{34}\:\mathcal{K}_{31}\:\mathcal{P}_{22}\:p^{2}_1
+\tfrac{1}{12}\:\mathcal{E}_{13}\:\mathcal{K}_{24}\:\mathcal{P}_{22}\:p^{2}_1
-\tfrac{1}{12}\:\mathcal{E}_{14}\:\mathcal{K}_{23}\:\mathcal{P}_{22}\:p^{2}_1
\vspace{.2cm} \nonumber \\ 
& \null
-\tfrac{1}{6}\:\mathcal{E}_{14}\:\mathcal{K}_{13}\:\mathcal{P}_{22}\:p^{2}_1
+\tfrac{1}{12}\:\mathcal{E}_{34}\:\mathcal{K}_{42}\:\mathcal{P}_{11}\:p^{2}_1
-\tfrac{1}{12}\:\mathcal{E}_{23}\:\mathcal{K}_{34}\:\mathcal{P}_{11}\:p^{2}_1
-\tfrac{1}{12}\:\mathcal{E}_{23}\:\mathcal{K}_{24}\:\mathcal{P}_{11}\:p^{2}_1
\vspace{.2cm} \nonumber \\ 
& \null
-\tfrac{1}{12}\:\mathcal{E}_{24}\:\mathcal{K}_{23}\:\mathcal{P}_{11}\:p^{2}_1
+\tfrac{1}{12}\:\mathcal{E}_{34}\:\mathcal{K}_{12}\:\mathcal{P}_{11}\:p^{2}_1
+\tfrac{1}{12}\:\mathcal{E}_{34}\:\mathcal{K}_{41}\:\mathcal{K}_{42}\:p^{2}_1
-\tfrac{1}{12}\:\mathcal{E}_{13}\:\mathcal{K}_{34}\:\mathcal{K}_{42}\:p^{2}_1
\vspace{.2cm} \nonumber \\ 
& \null
+\tfrac{1}{12}\:\mathcal{E}_{34}\:\mathcal{K}_{31}\:\mathcal{K}_{42}\:p^{2}_1
-\tfrac{1}{6}\:\mathcal{E}_{23}\:\mathcal{K}_{34}\:\mathcal{K}_{41}\:p^{2}_1
-\tfrac{1}{12}\:\mathcal{E}_{23}\:\mathcal{K}_{24}\:\mathcal{K}_{41}\:p^{2}_1
-\tfrac{1}{12}\:\mathcal{E}_{24}\:\mathcal{K}_{23}\:\mathcal{K}_{41}\:p^{2}_1
\vspace{.2cm} \nonumber \\ 
& \null
+\tfrac{1}{12}\:\mathcal{E}_{34}\:\mathcal{K}_{12}\:\mathcal{K}_{41}\:p^{2}_1
-\tfrac{1}{12}\:\mathcal{E}_{23}\:\mathcal{K}_{31}\:\mathcal{K}_{34}\:p^{2}_1
+\tfrac{1}{12}\:\mathcal{E}_{12}\:\mathcal{K}_{23}\:\mathcal{K}_{34}\:p^{2}_1
-\tfrac{1}{12}\:\mathcal{E}_{13}\:\mathcal{K}_{12}\:\mathcal{K}_{34}\:p^{2}_1
\vspace{.2cm} \nonumber \\ 
& \null
-\tfrac{1}{12}\:\mathcal{E}_{23}\:\mathcal{K}_{24}\:\mathcal{K}_{31}\:p^{2}_1
-\tfrac{1}{12}\:\mathcal{E}_{24}\:\mathcal{K}_{23}\:\mathcal{K}_{31}\:p^{2}_1
+\tfrac{1}{12}\:\mathcal{E}_{34}\:\mathcal{K}_{12}\:\mathcal{K}_{31}\:p^{2}_1
\Bigr]
+\mathrm{cyclic}.
\end{align}
%%%%%%%%%%%%%%
The notation `$+$ cyclic' indicates a sum over the three additional cyclic
permutations of indices, giving a total of four permutations $(1,2,3,4)$, $(2,3,4,1)$, $(3,4,1,2)$, and $(4,1,2,3)$ of the possible variables $\varepsilon_i, k_i, p_i, s\equiv(k_{1}+k_{2})^{2},$ and $t\equiv(k_{2}+k_{3})^{2}$.

The contribution from the gluon is the sum of a piece proportional to the scalar contribution and extra terms, denoted $n^{\text{(extra)}}_{1234;p}$. Explicitly,
%%%%%Gluon Loop%%%%%%
\begin{align}
\label{eq:N0D}
n^{\text{(gluon)}}_{1234;p}\: =\mathfrak{D}_{g}\:n^{\text{(scalar)}}_{1234;p}+n^{\text{(extra)}}_{1234;p},
\end{align}
where the proportionality factor, $\mathfrak{D}_{g}\equiv D-2$, is the number of gluonic states---i.e., on-shell degrees of freedom. 
The extra terms contribute the following:
%%%%%%%%%%%%%Extra Loop%%%%%%%%%%%%
\begin{align}
 \lefteqn{ \hskip 0 cm 
n^{\text{(extra)}}_{1234;p}\: =-i \Bigl[
\mathcal{E}_{12}\:\mathcal{E}_{34}\:p^{2}_1\:p^{2}_3
-\mathcal{E}_{13}\:\mathcal{E}_{24}\:p^{2}_1\:p^{2}_3
-\mathcal{E}_{12}\:\mathcal{E}_{34}\:p^{2}_1\:p^{2}_2
+\mathcal{E}_{13}\:\mathcal{E}_{24}\:p^{2}_1\:p^{2}_2
-\mathcal{E}_{14}\:\mathcal{E}_{23}\:p^{2}_1\:p^{2}_2}
\vspace{.2cm} \nonumber \\ 
& \null
+\mathcal{E}_{14}\:\mathcal{E}_{23}\left(p^{2}_1\right)^2
+4\:\mathcal{E}_{23}\:\mathcal{K}_{41}\:\mathcal{P}_{44}\:p^{2}_1
-4\:\mathcal{E}_{12}\:\mathcal{K}_{34}\:\mathcal{P}_{33}\:p^{2}_1
+4\:\mathcal{E}_{24}\:\mathcal{K}_{31}\:\mathcal{P}_{33}\:p^{2}_1
-4\:\mathcal{E}_{34}\:\mathcal{K}_{41}\:\mathcal{P}_{22}\:p^{2}_1
\vspace{.2cm} \nonumber \\ 
& \null
-4\:\mathcal{E}_{34}\:\mathcal{K}_{31}\:\mathcal{P}_{22}\:p^{2}_1
-4\:\mathcal{E}_{13}\:\mathcal{K}_{24}\:\mathcal{P}_{22}\:p^{2}_1
+4\:\mathcal{E}_{23}\:\mathcal{K}_{34}\:\mathcal{P}_{11}\:p^{2}_1
+4\:\mathcal{E}_{23}\:\mathcal{K}_{24}\:\mathcal{P}_{11}\:p^{2}_1
+\mathcal{E}_{12}\:\mathcal{E}_{34}\:p^{2}_2\:s
\vspace{.2cm} \nonumber \\ 
& \null
-\mathcal{E}_{13}\:\mathcal{E}_{24}\:p^{2}_2\:s
+\mathcal{E}_{14}\:\mathcal{E}_{23}\:p^{2}_2\:s
-\mathcal{E}_{14}\:\mathcal{E}_{23}\:p^{2}_1\:s
+2\:\mathcal{E}_{13}\:\mathcal{P}_{22}\:\mathcal{P}_{44}\:s
+2\:\mathcal{E}_{24}\:\mathcal{P}_{11}\:\mathcal{P}_{33}\:s
\vspace{.2cm} \nonumber \\ 
& \null
-4\:\mathcal{E}_{34}\:\mathcal{P}_{11}\:\mathcal{P}_{22}\:s
-2\:\mathcal{E}_{34}\:\mathcal{K}_{41}\:\mathcal{K}_{42}\:p^{2}_1
-2\:\mathcal{E}_{34}\:\mathcal{K}_{31}\:\mathcal{K}_{42}\:p^{2}_1
-2\:\mathcal{E}_{13}\:\mathcal{K}_{24}\:\mathcal{K}_{42}\:p^{2}_1
+2\:\mathcal{E}_{14}\:\mathcal{K}_{23}\:\mathcal{K}_{42}\:p^{2}_1
\vspace{.2cm} \nonumber \\ 
& \null
+6\:\mathcal{E}_{23}\:\mathcal{K}_{34}\:\mathcal{K}_{41}\:p^{2}_1
+4\:\mathcal{E}_{23}\:\mathcal{K}_{24}\:\mathcal{K}_{41}\:p^{2}_1
-2\:\mathcal{E}_{34}\:\mathcal{K}_{12}\:\mathcal{K}_{41}\:p^{2}_1
+2\:\mathcal{E}_{23}\:\mathcal{K}_{31}\:\mathcal{K}_{34}\:p^{2}_1
\vspace{.2cm} \nonumber \\ 
& \null
-2\:\mathcal{E}_{12}\:\mathcal{K}_{23}\:\mathcal{K}_{34}\:p^{2}_1
+2\:\mathcal{E}_{23}\:\mathcal{K}_{24}\:\mathcal{K}_{31}\:p^{2}_1
+2\:\mathcal{E}_{24}\:\mathcal{K}_{23}\:\mathcal{K}_{31}\:p^{2}_1
-2\:\mathcal{E}_{34}\:\mathcal{K}_{12}\:\mathcal{K}_{31}\:p^{2}_1
\vspace{.2cm} \nonumber \\ 
& \null
-2\:\mathcal{E}_{13}\:\mathcal{K}_{12}\:\mathcal{K}_{24}\:p^{2}_1
+2\:\mathcal{E}_{14}\:\mathcal{K}_{12}\:\mathcal{K}_{23}\:p^{2}_1
+4\:\mathcal{K}_{24}\:\mathcal{K}_{42}\:\mathcal{P}_{11}\:\mathcal{P}_{33}
-8\:\mathcal{K}_{23}\:\mathcal{K}_{34}\:\mathcal{P}_{11}\:\mathcal{P}_{22}
\vspace{.2cm} \nonumber \\ 
& \null
-8\:\mathcal{K}_{13}\:\mathcal{K}_{34}\:\mathcal{P}_{11}\:\mathcal{P}_{22}
-2\:\mathcal{E}_{34}\:\mathcal{K}_{41}\:\mathcal{P}_{22}\:s
+2\:\mathcal{E}_{13}\:\mathcal{K}_{34}\:\mathcal{P}_{22}\:s
+2\:\mathcal{E}_{13}\:\mathcal{K}_{24}\:\mathcal{P}_{22}\:s
-2\:\mathcal{E}_{14}\:\mathcal{K}_{23}\:\mathcal{P}_{22}\:s
\vspace{.2cm} \nonumber \\ 
& \null
-2\:\mathcal{E}_{34}\:\mathcal{K}_{42}\:\mathcal{P}_{11}\:s
-2\:\mathcal{E}_{23}\:\mathcal{K}_{34}\:\mathcal{P}_{11}\:s
-2\:\mathcal{E}_{23}\:\mathcal{K}_{24}\:\mathcal{P}_{11}\:s
+2\:\mathcal{E}_{24}\:\mathcal{K}_{23}\:\mathcal{P}_{11}\:s
-2\:\mathcal{E}_{34}\:\mathcal{K}_{12}\:\mathcal{P}_{11}\:s
\vspace{.2cm} \nonumber \\ 
& \null
-4\:\mathcal{K}_{23}\:\mathcal{K}_{34}\:\mathcal{K}_{42}\:\mathcal{P}_{11}
+4\:\mathcal{K}_{23}\:\mathcal{K}_{24}\:\mathcal{K}_{42}\:\mathcal{P}_{11}
+4\:\mathcal{K}_{13}\:\mathcal{K}_{24}\:\mathcal{K}_{42}\:\mathcal{P}_{11}
+4\:\mathcal{K}_{12}\:\mathcal{K}_{23}\:\mathcal{K}_{24}\:\mathcal{P}_{11}
\vspace{.2cm} \nonumber \\ 
& \null
+4\:\mathcal{K}_{12}\:\mathcal{K}_{13}\:\mathcal{K}_{24}\:\mathcal{P}_{11}
+\tfrac{1}{2}\:\mathcal{E}_{14}\:\mathcal{E}_{23}\:s^2
-2\:\mathcal{E}_{23}\:\mathcal{K}_{24}\:\mathcal{K}_{41}\:s
-2\:\mathcal{E}_{12}\:\mathcal{K}_{23}\:\mathcal{K}_{34}\:s
-2\:\mathcal{E}_{23}\:\mathcal{K}_{24}\:\mathcal{K}_{31}\:s
\vspace{.2cm} \nonumber \\ 
& \null
-2\:\mathcal{E}_{12}\:\mathcal{K}_{23}\:\mathcal{K}_{24}\:s
-4\:\mathcal{E}_{14}\:\mathcal{K}_{12}\:\mathcal{K}_{23}\:s
-2\:\mathcal{E}_{14}\:\mathcal{K}_{12}\:\mathcal{K}_{13}\:s
+\mathcal{K}_{13}\:\mathcal{K}_{24}\:\mathcal{K}_{31}\:\mathcal{K}_{42}
+2\:\mathcal{K}_{12}\:\mathcal{K}_{23}\:\mathcal{K}_{34}\:\mathcal{K}_{41}
\vspace{.2cm} \nonumber \\ 
& \null
+4\:\mathcal{K}_{12}\:\mathcal{K}_{23}\:\mathcal{K}_{31}\:\mathcal{K}_{34}
+2\:\mathcal{K}_{12}\:\mathcal{K}_{13}\:\mathcal{K}_{31}\:\mathcal{K}_{34}
+4\:\mathcal{K}_{12}\:\mathcal{K}_{23}\:\mathcal{K}_{24}\:\mathcal{K}_{31}
+4\:\mathcal{K}_{12}\:\mathcal{K}_{13}\:\mathcal{K}_{24}\:\mathcal{K}_{31}
\Bigr]
+\mathrm{cyclic}.
\end{align}
%%%%%%%%%%%%%%%

Finally, we give the contribution from the fermion loop:
%%%%%Fermion Loop%%%%%%
\begin{align}
\lefteqn{ \hskip 0 cm 
n^{\text{(fermion)}}_{1234;p}\: =-i\:\mathfrak{D}_{\!f}\:\Bigl[
-\tfrac{1}{12}\:\mathcal{E}_{12}\:\mathcal{E}_{34}\:p^{2}_1\:p^{2}_3
+\tfrac{1}{12}\:\mathcal{E}_{13}\:\mathcal{E}_{24}\:p^{2}_1\:p^{2}_3
-\tfrac{1}{8}\:\mathcal{E}_{14}\:\mathcal{E}_{23}\:p^{2}_1\:p^{2}_3
+\tfrac{1}{8}\:\mathcal{E}_{12}\:\mathcal{E}_{34}\:p^{2}_1\:p^{2}_2}
\vspace{.2cm} \nonumber \\ 
& \null
-\tfrac{1}{8}\:\mathcal{E}_{13}\:\mathcal{E}_{24}\:p^{2}_1\:p^{2}_2
+\tfrac{1}{8}\:\mathcal{E}_{14}\:\mathcal{E}_{23}\:p^{2}_1\:p^{2}_2
-\tfrac{1}{8}\:\mathcal{E}_{14}\:\mathcal{E}_{23}\left(p^{2}_1\right)^2
-\tfrac{1}{6}\:\mathcal{E}_{12}\:\mathcal{P}_{33}\:\mathcal{P}_{44}\:p^{2}_1
+\tfrac{1}{6}\:\mathcal{E}_{13}\:\mathcal{P}_{22}\:\mathcal{P}_{44}\:p^{2}_1
\vspace{.2cm} \nonumber \\ 
& \null
+\:\mathcal{E}_{14}\:\mathcal{P}_{22}\:\mathcal{P}_{33}\:p^{2}_1
+\tfrac{1}{6}\:\mathcal{E}_{24}\:\mathcal{P}_{11}\:\mathcal{P}_{33}\:p^{2}_1
-\tfrac{1}{6}\:\mathcal{E}_{34}\:\mathcal{P}_{11}\:\mathcal{P}_{22}\:p^{2}_1
-\:\mathcal{P}_{11}\:\mathcal{P}_{22}\:\mathcal{P}_{33}\:\mathcal{P}_{44}
+\tfrac{1}{12}\:\mathcal{E}_{13}\:\mathcal{K}_{42}\:\mathcal{P}_{44}\:p^{2}_1
\vspace{.2cm} \nonumber \\ 
& \null
-\tfrac{5}{12}\:\mathcal{E}_{23}\:\mathcal{K}_{41}\:\mathcal{P}_{44}\:p^{2}_1
-\tfrac{1}{12}\:\mathcal{E}_{12}\:\mathcal{K}_{23}\:\mathcal{P}_{44}\:p^{2}_1
+\tfrac{1}{12}\:\mathcal{E}_{13}\:\mathcal{K}_{12}\:\mathcal{P}_{44}\:p^{2}_1
-\tfrac{1}{12}\:\mathcal{E}_{14}\:\mathcal{K}_{42}\:\mathcal{P}_{33}\:p^{2}_1
\vspace{.2cm} \nonumber \\ 
& \null
+\tfrac{1}{12}\:\mathcal{E}_{24}\:\mathcal{K}_{41}\:\mathcal{P}_{33}\:p^{2}_1
+\tfrac{1}{4}\:\mathcal{E}_{12}\:\mathcal{K}_{34}\:\mathcal{P}_{33}\:p^{2}_1
-\tfrac{1}{3}\:\mathcal{E}_{24}\:\mathcal{K}_{31}\:\mathcal{P}_{33}\:p^{2}_1
-\tfrac{1}{12}\:\mathcal{E}_{12}\:\mathcal{K}_{24}\:\mathcal{P}_{33}\:p^{2}_1
\vspace{.2cm} \nonumber \\ 
& \null
+\tfrac{1}{12}\:\mathcal{E}_{14}\:\mathcal{K}_{12}\:\mathcal{P}_{33}\:p^{2}_1
+\tfrac{1}{4}\:\mathcal{E}_{34}\:\mathcal{K}_{41}\:\mathcal{P}_{22}\:p^{2}_1
+\tfrac{1}{12}\:\mathcal{E}_{13}\:\mathcal{K}_{34}\:\mathcal{P}_{22}\:p^{2}_1
+\tfrac{1}{3}\:\mathcal{E}_{34}\:\mathcal{K}_{31}\:\mathcal{P}_{22}\:p^{2}_1
\vspace{.2cm} \nonumber \\ 
& \null
+\tfrac{5}{12}\:\mathcal{E}_{13}\:\mathcal{K}_{24}\:\mathcal{P}_{22}\:p^{2}_1
+\tfrac{1}{12}\:\mathcal{E}_{14}\:\mathcal{K}_{23}\:\mathcal{P}_{22}\:p^{2}_1
+\tfrac{1}{6}\:\mathcal{E}_{14}\:\mathcal{K}_{13}\:\mathcal{P}_{22}\:p^{2}_1
-\tfrac{1}{12}\:\mathcal{E}_{34}\:\mathcal{K}_{42}\:\mathcal{P}_{11}\:p^{2}_1
\vspace{.2cm} \nonumber \\ 
& \null
-\tfrac{5}{12}\:\mathcal{E}_{23}\:\mathcal{K}_{34}\:\mathcal{P}_{11}\:p^{2}_1
-\tfrac{5}{12}\:\mathcal{E}_{23}\:\mathcal{K}_{24}\:\mathcal{P}_{11}\:p^{2}_1
+\tfrac{1}{12}\:\mathcal{E}_{24}\:\mathcal{K}_{23}\:\mathcal{P}_{11}\:p^{2}_1
-\tfrac{1}{12}\:\mathcal{E}_{34}\:\mathcal{K}_{12}\:\mathcal{P}_{11}\:p^{2}_1
\vspace{.2cm} \nonumber \\ 
& \null
-\tfrac{1}{8}\:\mathcal{E}_{12}\:\mathcal{E}_{34}\:p^{2}_2\:s
+\tfrac{1}{8}\:\mathcal{E}_{13}\:\mathcal{E}_{24}\:p^{2}_2\:s
-\tfrac{1}{8}\:\mathcal{E}_{14}\:\mathcal{E}_{23}\:p^{2}_2\:s
+\tfrac{1}{8}\:\mathcal{E}_{14}\:\mathcal{E}_{23}\:p^{2}_1\:s
-\tfrac{1}{4}\:\mathcal{E}_{13}\:\mathcal{P}_{22}\:\mathcal{P}_{44}\:s
\vspace{.2cm} \nonumber \\ 
& \null
-\tfrac{1}{4}\:\mathcal{E}_{24}\:\mathcal{P}_{11}\:\mathcal{P}_{33}\:s
+\tfrac{1}{2}\:\mathcal{E}_{34}\:\mathcal{P}_{11}\:\mathcal{P}_{22}\:s
+\tfrac{1}{6}\:\mathcal{E}_{34}\:\mathcal{K}_{41}\:\mathcal{K}_{42}\:p^{2}_1
+\tfrac{1}{12}\:\mathcal{E}_{13}\:\mathcal{K}_{34}\:\mathcal{K}_{42}\:p^{2}_1
+\tfrac{1}{6}\:\mathcal{E}_{34}\:\mathcal{K}_{31}\:\mathcal{K}_{42}\:p^{2}_1
\vspace{.2cm} \nonumber \\ 
& \null
+\tfrac{1}{4}\:\mathcal{E}_{13}\:\mathcal{K}_{24}\:\mathcal{K}_{42}\:p^{2}_1
-\tfrac{1}{4}\:\mathcal{E}_{14}\:\mathcal{K}_{23}\:\mathcal{K}_{42}\:p^{2}_1
-\tfrac{7}{12}\:\mathcal{E}_{23}\:\mathcal{K}_{34}\:\mathcal{K}_{41}\:p^{2}_1
-\tfrac{5}{12}\:\mathcal{E}_{23}\:\mathcal{K}_{24}\:\mathcal{K}_{41}\:p^{2}_1
\vspace{.2cm} \nonumber \\ 
& \null
+\tfrac{1}{12}\:\mathcal{E}_{24}\:\mathcal{K}_{23}\:\mathcal{K}_{41}\:p^{2}_1
+\tfrac{1}{6}\:\mathcal{E}_{34}\:\mathcal{K}_{12}\:\mathcal{K}_{41}\:p^{2}_1
-\tfrac{1}{6}\:\mathcal{E}_{23}\:\mathcal{K}_{31}\:\mathcal{K}_{34}\:p^{2}_1
+\tfrac{1}{6}\:\mathcal{E}_{12}\:\mathcal{K}_{23}\:\mathcal{K}_{34}\:p^{2}_1
\vspace{.2cm} \nonumber \\ 
& \null
+\tfrac{1}{12}\:\mathcal{E}_{13}\:\mathcal{K}_{12}\:\mathcal{K}_{34}\:p^{2}_1
-\tfrac{1}{6}\:\mathcal{E}_{23}\:\mathcal{K}_{24}\:\mathcal{K}_{31}\:p^{2}_1
-\tfrac{1}{6}\:\mathcal{E}_{24}\:\mathcal{K}_{23}\:\mathcal{K}_{31}\:p^{2}_1
+\tfrac{1}{6}\:\mathcal{E}_{34}\:\mathcal{K}_{12}\:\mathcal{K}_{31}\:p^{2}_1
\vspace{.2cm} \nonumber \\ 
& \null
+\tfrac{1}{4}\:\mathcal{E}_{13}\:\mathcal{K}_{12}\:\mathcal{K}_{24}\:p^{2}_1
-\tfrac{1}{4}\:\mathcal{E}_{14}\:\mathcal{K}_{12}\:\mathcal{K}_{23}\:p^{2}_1
-\tfrac{1}{2}\:\mathcal{K}_{24}\:\mathcal{K}_{42}\:\mathcal{P}_{11}\:\mathcal{P}_{33}
+\:\mathcal{K}_{23}\:\mathcal{K}_{34}\:\mathcal{P}_{11}\:\mathcal{P}_{22}
\vspace{.2cm} \nonumber \\ 
& \null
+\:\mathcal{K}_{13}\:\mathcal{K}_{34}\:\mathcal{P}_{11}\:\mathcal{P}_{22}
+\tfrac{1}{4}\:\mathcal{E}_{34}\:\mathcal{K}_{41}\:\mathcal{P}_{22}\:s
-\tfrac{1}{4}\:\mathcal{E}_{13}\:\mathcal{K}_{34}\:\mathcal{P}_{22}\:s
-\tfrac{1}{4}\:\mathcal{E}_{13}\:\mathcal{K}_{24}\:\mathcal{P}_{22}\:s
+\tfrac{1}{4}\:\mathcal{E}_{14}\:\mathcal{K}_{23}\:\mathcal{P}_{22}\:s
\vspace{.2cm} \nonumber \\ 
& \null
+\tfrac{1}{4}\:\mathcal{E}_{34}\:\mathcal{K}_{42}\:\mathcal{P}_{11}\:s
+\tfrac{1}{4}\:\mathcal{E}_{23}\:\mathcal{K}_{34}\:\mathcal{P}_{11}\:s
+\tfrac{1}{4}\:\mathcal{E}_{23}\:\mathcal{K}_{24}\:\mathcal{P}_{11}\:s
-\tfrac{1}{4}\:\mathcal{E}_{24}\:\mathcal{K}_{23}\:\mathcal{P}_{11}\:s
+\tfrac{1}{4}\:\mathcal{E}_{34}\:\mathcal{K}_{12}\:\mathcal{P}_{11}\:s
\vspace{.2cm} \nonumber \\ 
& \null
+\tfrac{1}{2}\:\mathcal{K}_{23}\:\mathcal{K}_{34}\:\mathcal{K}_{42}\:\mathcal{P}_{11}
-\tfrac{1}{2}\:\mathcal{K}_{23}\:\mathcal{K}_{24}\:\mathcal{K}_{42}\:\mathcal{P}_{11}
-\tfrac{1}{2}\:\mathcal{K}_{13}\:\mathcal{K}_{24}\:\mathcal{K}_{42}\:\mathcal{P}_{11}
-\tfrac{1}{2}\:\mathcal{K}_{12}\:\mathcal{K}_{23}\:\mathcal{K}_{24}\:\mathcal{P}_{11}
\vspace{.2cm} \nonumber \\ 
& \null
-\tfrac{1}{2}\:\mathcal{K}_{12}\:\mathcal{K}_{13}\:\mathcal{K}_{24}\:\mathcal{P}_{11}
-\tfrac{1}{32}\:\mathcal{E}_{13}\:\mathcal{E}_{24}\:s\:t
+\tfrac{1}{16}\:\mathcal{E}_{14}\:\mathcal{E}_{23}\:s\:t
-\tfrac{1}{8}\:\mathcal{E}_{13}\:\mathcal{K}_{24}\:\mathcal{K}_{42}\:s
+\tfrac{1}{4}\:\mathcal{E}_{23}\:\mathcal{K}_{24}\:\mathcal{K}_{41}\:s
\vspace{.2cm} \nonumber \\ 
& \null
-\tfrac{1}{8}\:\mathcal{E}_{24}\:\mathcal{K}_{23}\:\mathcal{K}_{41}\:s
-\tfrac{1}{8}\:\mathcal{E}_{13}\:\mathcal{K}_{12}\:\mathcal{K}_{34}\:s
-\tfrac{1}{4}\:\mathcal{E}_{13}\:\mathcal{K}_{12}\:\mathcal{K}_{24}\:s
+\tfrac{1}{4}\:\mathcal{E}_{14}\:\mathcal{K}_{12}\:\mathcal{K}_{23}\:s
-\tfrac{1}{8}\:\mathcal{K}_{13}\:\mathcal{K}_{24}\:\mathcal{K}_{31}\:\mathcal{K}_{42}
\vspace{.2cm} \nonumber \\ 
& \null
-\tfrac{1}{4}\:\mathcal{K}_{12}\:\mathcal{K}_{23}\:\mathcal{K}_{34}\:\mathcal{K}_{41}
-\tfrac{1}{2}\:\mathcal{K}_{12}\:\mathcal{K}_{23}\:\mathcal{K}_{31}\:\mathcal{K}_{34}
-\tfrac{1}{4}\:\mathcal{K}_{12}\:\mathcal{K}_{13}\:\mathcal{K}_{31}\:\mathcal{K}_{34}
-\tfrac{1}{2}\:\mathcal{K}_{12}\:\mathcal{K}_{23}\:\mathcal{K}_{24}\:\mathcal{K}_{31}
\vspace{.2cm} \nonumber \\ 
& \null
-\tfrac{1}{2}\:\mathcal{K}_{12}\:\mathcal{K}_{13}\:\mathcal{K}_{24}\:\mathcal{K}_{31}
\Bigr]
+\mathrm{cyclic}.
\label{fermion}
\end{align}
%%%%%%%%%%%%%%% 
Here, there is also a well-known proportionality
factor, $\mathfrak{D}_{\!f}$, that denotes the number of
states---i.e., on-shell degrees of freedom---of each 
fermion. The
minimal spinor type corresponding to each spacetime dimension is
provided in \tab{tab:minSpinor} along with its number of
states (see Ref.~\cite{Freedman}).
%%%%%%%%%%%Table%%%%%%%%%%%%%%%%%%
\begin{table}
  \begin{tabular}{ c || c | c }
  Dimension ($D$) &\phantom{--------}Minimal Spinor Type\phantom{--------}& \# of States ($\mathfrak{D}_{\!f}$)\\
    \hline \hline
 
 $3$ &
    	 Majorana&
	 $1$\\ \hline
	 
 $4$ &
    	 Majorana&
	 $2$\\ \hline
	  
 $5$ &
    	Dirac&
	 $4$\\ \hline 
	 
 $6$ &
    	Weyl&
	 $4$\\ \hline
	  
 $7$ &
    	 Dirac&
	 $8$\\ \hline
	  
 $8$ &
    	 Pseudo-Majorana&
	 $8$\\ \hline
	  
 $9$ &
    	 Pseudo-Majorana&
	 $8$\\ \hline
	  
 $10$ &
    	  Pseudo-Majorana and Weyl&
	 $8$\\ \hline
	  
 $11$ &
    	 Majorana&
	 $16$\\ \hline
  \end{tabular}
  \caption{The number of states in minimal spinors, dependent on dimension. We note that in $D=5,6,$ and $7$, a symplectic Majorana condition can be applied among an even number of spinors. We ignore this condition here.}
  \label{tab:minSpinor}
\end{table}
%%%%%%%%%%%%%%%%%%%%%%%%%%%%%

It is now simple to obtain BCJ numerators with four-dimensional 
external states that we use to compare to earlier work in \sect{sec:NumComp}:
\begin{subequations}
\label{eqn:numComp}
\begin{alignat}{13}
&n^{\mathcal{N}=4}_{1234;p}&&=&&\left.n^{\text{(gluon)}}_{1234;p}\right|_{\mathfrak{D}_{g}=2}&&+4 &&\left.n^{\text{(fermion)}}_{1234;p}\right|_{\mathfrak{D}_{\!f}=2}&&+6 \:&&n^{\text{(scalar)}}_{1234;p},
\\[.25cm]
&n^{\mathcal{N}=1\text{(chiral)}}_{1234;p}&&=&&&&&&\left.n^{\text{(fermion)}}_{1234;p}\right|_{\mathfrak{D}_{\!f}=2}&&+2\:&&n^{\text{(scalar)}}_{1234;p},
\\[.25cm]
&n^{\mathcal{N}=0}_{1234;p}&&=&&\left.n^{\text{(gluon)}}_{1234;p}\right|_{\mathfrak{D}_{g}=2}.
\label{eq:N0num}
\end{alignat}
\end{subequations}
To explicitly see the simplification due to supersymmetry, we provide the $\mathcal{N}=4$ and $\mathcal{N}=1$ (chiral) box numerators in \app{sec:num4}. In \sect{sec:NumComp}, we compare the numerators of Eqs.~(\ref{eqn:numComp}) to the $\mathcal{N}=4$ numerators of Ref.~\cite{N4}, the $\mathcal{N}=1$ (chiral) MHV numerators of Ref.~\cite{JJRadu}, and the $\mathcal{N}=0$ all-plus-helicity numerators of Ref.~\cite{BernMorgan}. But first, we discuss how to put these formal-polarization expressions into a helicity basis in the next section.

Before we proceed, we clarify a few points related to our inclusion of
matter. First, we reiterate that using nonsupersymmetric field content
in \eqn{eq:numDecomp} still yields valid BCJ numerators, but
numerators with internal bubbles or bubbles on external legs no longer
vanish and the loop momentum power counting is not improved. Second,
our $\mathcal{N}=0$ numerators---generated by the box numerator of
\eqn{eq:N0D}---differ from those presented in Ref.~\cite{nonSUSYBCJ}
because of our restrictions on the internal bubble numerators. Third,
we emphasize that we are only considering numerators with external
gluons. Amplitudes with arbitrary field content on the
external legs do not always straightforwardly allow a BCJ
representation.  In particular, consider two different flavors of
scalars minimally coupled to nonsupersymmetric YM theory. We notice at
tree level that four external scalars of two different flavors can
only scatter in one channel. Thus, we cannot satisfy color-kinematics
duality because the numerator Jacobi relations relate three different
channels. A remedy in this situation is to add a four-point contact
interaction that mixes the different flavors~\cite{Matter, Vera,
  Anomaly}. This is the interaction that arises when nonsupersymmetric
pure YM theory is dimensionally reduced from six dimensions to four
dimensions. This has been studied in some detail in the context of
multi-Regge kinematics in Ref.~\cite{Vera}. We expect similar
properties for external fermions. Namely, we expect that the duality
works straightforwardly with only one flavor\footnote{We note the
restrictions of Ref.~\cite{Orbifolds}: The four-fermion tree amplitude,
$A_{4}^{\text{tree}}(1^{\psi},2^{\psi},3^{\psi},4^{\psi})$, can only
satisfy color-kinematics duality in $D=3,4,6,10$. We thank Radu Roiban 
for bringing this to our attention.}; however, for multiple
flavors, we anticipate the need for flavor-mixing Yukawa
interactions. Such interactions are seen in $\mathcal{N}=2$ sYM theory
in four dimensions. ($\mathcal{N}=2$ sYM theory in four dimensions can
be constructed by dimensionally reducing six-dimensional
$\mathcal{N}=1$ (vector) sYM theory, which has only one fermion
flavor.) There are no such issues for our one-loop amplitudes with
external gluons and multiple flavors of matter in the loop; each
diagram can only have one flavor circulating in the loop at a
time. However, this is not the case for higher loop orders.

Finally, we mention that our BCJ numerators can be used in
\eqn{DoubleCopy} to calculate amplitudes in gravity theories. 
For our amplitudes, the external
states can consist of gravitons, antisymmetric tensors, and dilatons,
as discussed in Ref.~\cite{nonSUSYBCJ}. As an example of
field content in the loop, we note that the product of our gluon box
numerator and fermion box numerator,
$n^{\text{(gluon)}}_{1234;p}\times
n^{\text{(fermion)}}_{1234;p}$, gives the box numerator for a
gravitino and fermion circulating in the loop. This agrees with simple
state counting. The tensor product of the gluon and fermion states
yields $\mathfrak{D}_{g}\times\mathfrak{D}_{\!f}$ states. Likewise,
the total number of states of a gravitino and a fermion is
$(\mathfrak{D}_{g}-1)\mathfrak{D}_{\!f}+\mathfrak{D}_{\!f}=\mathfrak{D}_{g}\times\mathfrak{D}_{\!f}$.

Plain-text, computer-readable versions of the full expressions 
for the numerators can be found online~\cite{AncFile}.
%%%%%%%%%%%%%%%%%%%%%%%%%%%%%%%%%%%%%%%%%%%%%%%

%%%%%%%%%%%%%%%%%%%%%%%%%%%%%%%%%%%%%%%%%%%%%%%%
\section{Polarization Vectors in a Momentum Basis}
\label{sec:MomBasis}
To compare our results to existing literature, we consider our numerators in specific four-dimensional helicity configurations. (Because we use dimensional regularization, the loop momentum is in $D=4-2\varepsilon$.) We do this by putting the formal polarization vectors into a momentum basis, as in Ref.~\cite{epMomBasis}. Because we are dealing with four-point amplitudes in four dimensions, the momentum basis consists of three independent external momentum vectors and an orthogonal dual vector. The dual vector is formed by contracting three independent external momentum vectors with the four-dimensional Levi-Civita symbol:
\begin{align}
\label{eq:v}
	v^{\mu} \equiv \epsilon(\mu,k_{1},k_{2},k_{3})\equiv \epsilon^{\mu\alpha\beta\gamma}k_{1\alpha}k_{2\beta}k_{3\gamma}.
 \end{align}

We take care to preserve the phase factors associated with the polarization vectors. Phase factors arise naturally in the spinor-helicity formalism via
\begin{align}
\left\langle i j\right\rangle = \sqrt{\left| s_{ij}\right|}e^{i\phi_{ij}},
\hspace{15mm}
\left[ i j\right] = \sqrt{\left| s_{ij}\right|}e^{-i(\phi_{ij}+\pi)},
\end{align}
where particles $i$ and $j$ have positive energy, i.e., $k_{i}^{0}>0$ and $k_{j}^{0}>0$ (cf. Ref.~\cite{Dixon}). Also, we define $s_{ij}\equiv(k_{i}+k_{j})^{2}$, so $s\equiv s_{12}=s_{34}$, $t\equiv s_{23}=s_{14}$, and $u\equiv s_{13}=s_{24}=-s-t$ are the standard Mandelstam variables. By antisymmetry of the spinor products and momentum conservation,
\begin{align}
\label{eq:phi}
	\phi_{ji} = \phi_{ij}+\pi,
\hspace{10mm}
	\phi_{24} = -\phi_{13}+\phi_{14}+\phi_{23}+\pi,
\hspace{10mm}
	\phi_{34} = -\phi_{12}+\phi_{14}+\phi_{23}.
 \end{align}
The combination of phase factors that appear in our calculations are immediately identified with spinor-helicity expressions. For reference, we list the four-point combinations that arise:
\begin{align}
\begin{array}{rcccl}
	e^{-2 i (\phi_{14}+\phi_{23})} &=& \dfrac{\left[ 12 \right]\left[ 34 \right]}{\left< 12 \right>\left< 34 \right>} &\sim& \mathrm{helicity:} ++++, \vspace{.2cm} \\[.5cm]
	e^{2 i (\phi_{12}+\phi_{13}-\phi_{14}-2\phi_{23})} &=& -\dfrac{s t}{u} \dfrac{\left[ 24 \right]^{2}}{\left[ 12 \right]\left< 23 \right>\left< 34 \right>\left[ 41 \right]} &\sim& \mathrm{helicity:} -+++,\vspace{.2cm}\\[.5cm]
	e^{2 i (2\phi_{12}-\phi_{14}-\phi_{23})} &=& -\dfrac{t}{s}\dfrac{\left< 12 \right>^{4}}{\left< 12 \right>\left< 23 \right>\left< 34 \right>\left< 41 \right>} &\sim& \mathrm{helicity:} --++.
 \end{array}
\end{align}

Of course, some care is needed when using dimensional
regularization. The external momenta, $k^{\mu}_{i}$; formal
polarization vectors, $\varepsilon^{\mu}_{i}$; and dual vector,
$v^{\mu}$, are still four-dimensional objects. However, the loop
momentum, call it $\tilde{l}^{\mu}$, is in $4-2\varepsilon$
dimensions. We denote the four-dimensional components of the loop
momentum as $l^{\mu}$ and the spacelike ($-2\varepsilon$)-component as
$l_{[-2\varepsilon]}^{\mu}$. The $-2\varepsilon$ dimensions only affect the $\tilde{l}^{2}$ inner product. Specifically,
\begin{align}
\label{eq:mu}
\tilde{l}\cdot k_{i} = l\cdot k_{i}, \hspace{15mm}
\tilde{l}\cdot v = l\cdot v, \hspace{15mm}
\tilde{l}^{2}= l^{2} - l_{[-2\varepsilon]}^{2}\equiv l^{2}-\mu^{2},
\end{align}
noting that we use the mostly-minus metric convention. The $\mu^{2}\equiv l^{2}_{[-2 \varepsilon]}$ can be treated as an effective mass of the loop momentum. (For the importance of the $\mu^{2}$ in loop calculations, we direct the reader to Ref.~\cite{Badger}.)  

Using identities of the Levi-Civita symbol, we find the following properties of the dual vector:
\begin{align}
 	v^{2} &= -\tfrac{1}{4}stu,
\\[.25cm]
	(l\cdot v)^{2} &= -\tfrac{1}{4}\left[t \tau_{51}(t\tau_{51}-2u\tau_{52})+\text{(cyclic permutations of 1,2,3)}+stu(\tilde{\tau}_{55}+\mu^{2})\right] ,
\end{align}
where we define
\begin{align}
\label{eq:tau}
\tau_{5i}\equiv l\cdot k_{i}, \hspace{2cm} \tilde{\tau}_{55}\equiv \tilde{l}^{2}.
\end{align}
Because the propagators in our amplitudes contain the $(4-2\varepsilon)$-dimensional $\tilde{l}^{2}$'s, we write the $l^{2}$ term in $(l\cdot v)^{2}$ as $\tilde{l}^{2}+\mu^{2}$. This is the only vehicle through which $\mu^{2}$ terms arise in our expressions.

Now, we put the polarization vectors into a momentum basis. We define a four-dimensional representation of the polarization vector corresponding to an external leg with momentum $k_{i}$ by
\begin{align}
\varepsilon^{\mu}_{h_{i}}(i;j_{1},j_{2})&\equiv \mathcal{P}_{h_{i}}(i;j_{1},j_{2}) \sqrt{\frac{2}{s_{ij_{1}}s_{ij_{2}}s_{j_{1}j_{2}}}}\:\times \nonumber \\
&\hspace{.75cm}\left[\left(k_{j_{1}}\cdot k_{j_{2}}\right)k_{i}^{\mu}+\left(k_{i}\cdot k_{j_{2}}\right)k_{j_{1}}^{\mu}-\left(k_{i}\cdot k_{j_{1}}\right)k_{j_{2}}^{\mu}+i\:h_{i}\:\epsilon(\mu,k_{i},k_{j_{1}},k_{j_{2}})\right].
\label{eq:epDef}
\end{align}
The arguments $i$, $j_{1}$, $j_{2}$ correspond to the external momenta $k_{i}$, $k_{j_{1}}$, $k_{j_{2}}$, where $k_{j_{1}}$ and $k_{j_{2}}$ are the reference momenta. $\mu$ is a free Lorentz index and $h_{i}=\pm 1$ defines the helicity state, which we sometimes simply denote as $h_{i}=\pm$. $\mathcal{P}_{h_{i}}(i;j_{1},j_{2})$ is a phase factor that we determine in \app{sec:phase} to be
\begin{align}
\mathcal{P}_{h_{i}}(i;j_{1},j_{2})=-e^{-i h_{i} \left(\phi_{ij_{2}}-\phi_{j_{1}j_{2}}+\phi_{i j_{1}}\right)}.
\end{align}
The coefficients of the basis vectors were determined by demanding the following:
\begin{equation}
\begin{gathered}
k_{i}\cdot\varepsilon_{h_{i}}(i;j_{1},j_{2})=0,\hspace{8mm}
k_{j_{1}}\cdot\varepsilon_{h_{i}}(i;j_{1},j_{2})=0,\hspace{8mm}
\varepsilon^{*}_{\pm}(i;j_{1},j_{2})=\varepsilon_{\mp}(i;j_{1},j_{2}),\\[.5cm]
\varepsilon_{h_{i}}(i;j_{1},j_{2})\cdot\varepsilon_{h_{i}}(i;j_{1},j_{2})=0,\hspace{12mm}
\varepsilon_{h_{i}}(i;j_{1},j_{2})\cdot\varepsilon^{*}_{h_{i}}(i;j_{1},j_{2})=-1.
\end{gathered}
\label{eqn:epProp}
\end{equation}
Note that $i\neq j_{1}\neq j_{2}$; otherwise, $\epsilon(\mu,k_{i},k_{j_{1}},k_{j_{2}})=0$, and Eqs.~(\ref{eqn:epProp}) cannot be satisfied. This implies that $s_{ij_{1}}s_{ij_{2}}s_{j_{1}j_{2}}=stu$. 
 
Choosing different reference momenta in \eqn{eq:epDef} changes the expression by at most a gauge shift. For example,
\begin{subequations}
\begin{align}
 	\varepsilon_{h_{1}}^{\mu}(1;3,4)&=\varepsilon_{h_{1}}^{\mu}(1;2,3)-\left(\mathcal{P}_{h_{1}}(1;2,3)\sqrt{\tfrac{2}{stu}}\: t\right) k_{1}^{\mu},\\[.2cm]
	\varepsilon_{h_{3}}^{\mu}(3;4,1)&=\varepsilon_{h_{3}}^{\mu}(3;1,2)+\left(\mathcal{P}_{h_{3}}(3;1,2)\sqrt{\tfrac{2}{stu}}\: t\right) k_{3}^{\mu},
\end{align}
\end{subequations}
where we enforce momentum conservation. We choose the following reference momenta that simplify our $\mathcal{N}=1$ (chiral) result, which we present later:
\begingroup
\addtolength{\jot}{.1cm}
\begin{subequations}\label{eq:rep}
\begin{align}
 	\varepsilon_{1}^{\mu} & \to  \varepsilon_{h_{1}}^{\mu}(1;3,4),\\
	\varepsilon_{2}^{\mu} & \to  \varepsilon_{h_{2}}^{\mu}(2;3,1),\\
	\varepsilon_{3}^{\mu} & \to  \varepsilon_{h_{3}}^{\mu}(3;4,1), \\
	\varepsilon_{4}^{\mu} & \to  \varepsilon_{h_{4}}^{\mu}(4;3,1).
\end{align}
\end{subequations}
\endgroup
For consistency, we use this choice throughout the remainder of the paper. Other gauge choices do not affect the $\mathcal{N}=4$ result, and the $\mathcal{N}=0$ expression will be equally lengthy regardless of reference momenta choices. With this representation, we tabulate the inner products $\varepsilon_{i}\cdot\varepsilon_{j}$ in \tab{tab:ee1}, where we suppress the phase factors, and $\varepsilon_{i}\cdot\varepsilon_{j}^{*}$ in \tab{tab:ee2}. Also, we suppress the phase factor along with $\sqrt{2/(stu)}$ and list $\varepsilon_{i}\cdot k_{j}$ in \tab{tab:ke}.
%%%%%%%%%%%%Table%%%%%%%%%%%%%%%%%
\begin{table}
\bgroup
\def\arraystretch{1.5}
  \begin{tabular}{ c || c | c | c | c}
 $\varepsilon_{i}\cdot\varepsilon_{j}$ & $\varepsilon_{1}$ & $\varepsilon_{2}$ & $\varepsilon_{3}$ &$\varepsilon_{4}$\\
    \hline \hline
 
   $\varepsilon_{1}$ &
    	 $0$&
	 $-\tfrac{1}{2}(1-h_{1}h_{2})$&
	 $\tfrac{1}{2}(1+h_{1}h_{3})$&
	 $\tfrac{1}{2}(1-h_{1}h_{4})$ \\ \hline
   
    $\varepsilon_{2}$ & 
    	 $-\tfrac{1}{2}(1-h_{1}h_{2})$&
	 $0$&
	 $\tfrac{1}{2}(1+h_{2}h_{3})$&
	 $\tfrac{1}{2}(1-h_{2}h_{4})$ \\ \hline
	 
    $\varepsilon_{3}$ & 
    	 $\tfrac{1}{2}(1+h_{1}h_{3})$&
	 $\tfrac{1}{2}(1+h_{2}h_{3})$&
	 $0$&
	 $-\tfrac{1}{2}(1+h_{3}h_{4})$ \\ \hline
  
    $\varepsilon_{4}$ &
    	 $\tfrac{1}{2}(1-h_{1}h_{4})$&
	 $\tfrac{1}{2}(1-h_{2}h_{4})$&
	 $-\tfrac{1}{2}(1+h_{3}h_{4})$&
	 $0$ \\ \hline
  \end{tabular}
  \egroup
  \caption{The inner product $\varepsilon_{i}\cdot\varepsilon_{j}$ in the representation given by Eqs.~(\ref{eq:rep}) with phase factors suppressed. $h_{i}=\pm 1$ corresponds to the helicity of leg $i$.}
\label{tab:ee1}
\end{table}
%%%%%%%%%%%%%%%%%%%%%%%%%%%%%%%%
%%%%%%%%%%%%Table%%%%%%%%%%%%%%%%%
\begin{table}
\bgroup
\def\arraystretch{1.5}
  \begin{tabular}{ c || c | c | c | c}
 $\varepsilon_{i}\cdot\varepsilon^{*}_{j}$ & $\varepsilon^{*}_{1}$ & $\varepsilon^{*}_{2}$ & $\varepsilon^{*}_{3}$ &$\varepsilon^{*}_{4}$\\
    \hline \hline
   $\varepsilon_{1}$ &
    	 $-1$&
	 $-\tfrac{1}{2}(1+h_{1}h_{2})$&
	 $\tfrac{1}{2}(1-h_{1}h_{3})$&
	 $\tfrac{1}{2}(1+h_{1}h_{4})$ \\ \hline
   
    $\varepsilon_{2}$ & 
    	 $-\tfrac{1}{2}(1+h_{1}h_{2})$&
	 $-1$&
	 $\tfrac{1}{2}(1-h_{2}h_{3})$&
	 $\tfrac{1}{2}(1+h_{2}h_{4})$ \\ \hline
	 
    $\varepsilon_{3}$ & 
    	 $\tfrac{1}{2}(1-h_{1}h_{3})$&
	 $\tfrac{1}{2}(1-h_{2}h_{3})$&
	 $-1$&
	 $-\tfrac{1}{2}(1-h_{3}h_{4})$ \\ \hline
  
    $\varepsilon_{4}$ &
    	 $\tfrac{1}{2}(1+h_{1}h_{4})$&
	 $\tfrac{1}{2}(1+h_{2}h_{4})$&
	 $-\tfrac{1}{2}(1-h_{3}h_{4})$&
	 $-1$ \\ \hline
  \end{tabular}
  \egroup
    \caption{The inner product $\varepsilon_{i}\cdot\varepsilon^{*}_{j}$ in the representation given by Eqs.~(\ref{eq:rep}). $h_{i}=\pm 1$ corresponds to the helicity of leg $i$}
\label{tab:ee2}
\end{table}
%%%%%%%%%%%%%%%%%%%%%%%%%%%%%%%%
%%%%%%%%%%%%Table%%%%%%%%%%%%%%%%%
\begin{table}
\bgroup
\def\arraystretch{1.5}
  \begin{tabular}{ c || c | c | c | c}
 $\varepsilon_{i}\cdot k_{j}$ &$\phantom{-}k_{1}\phantom{-}$&$\phantom{-}k_{2}\phantom{-}$&$\phantom{-}k_{3}\phantom{-}$ &$\phantom{-}k_{4}\phantom{-}$\\
    \hline \hline
   $\varepsilon_{1}$ &
    	 $0$&
	 $-\tfrac{1}{2}st$&
	 $0$&
	 $\tfrac{1}{2}st$ \\ \hline
    $\varepsilon_{2}$ & 
    	 $\tfrac{1}{2}su$&
	 $0$&
	 $0$&
	 $-\tfrac{1}{2}su$ \\ \hline
    $\varepsilon_{3}$ & 
    	 $\tfrac{1}{2}tu$&
	 $-\tfrac{1}{2}tu$&
	 $0$&
	 $0$ \\ \hline
    $\varepsilon_{4}$ &
    	 $\tfrac{1}{2}tu$&
	 $-\tfrac{1}{2}su$&
	 $0$&
	 $0$ \\ \hline
  \end{tabular}
  \egroup
   \caption{The inner product $\varepsilon_{i}\cdot k_{j}$ in the representation given by Eqs.~(\ref{eq:rep}), suppressing the phase factor and $\sqrt{2/(stu)}$.  }
\label{tab:ke}
\end{table}
%%%%%%%%%%%%%%%%%%%%%%%%%%%%%%%%

We draw attention to the fact that putting the polarization vectors into a momentum basis using the prescription above introduces a degree of non-locality. Each expression in a four-point numerator will have the non-local factor $4/(stu)^{2}$ since $\varepsilon_{1}$, $\varepsilon_{2}$, $\varepsilon_{3}$, and $\varepsilon_{4}$ are present in each term. This is the same degree of non-locality that is present in the $\mathcal{N}=1$ (chiral) numerators of Ref.~\cite{JJRadu}. In addition, the relabeling symmetries of the formal-polarization numerators in \sect{sec:formal} are, in general, lost once the polarization vectors are put into a momentum basis. Thus, instead of one box numerator, there are now three unrelated by relabeling.
%%%%%%%%%%%%%%%%%%%%%%%%%%%%%%%%%%%%%%%%%%%%%%%

\section{BCJ Numerator Comparisons}
\label{sec:NumComp}
Here, we compare our BCJ numerators to existing representations in literature. Even if two sets of BCJ numerators satisfy the color-kinematics duality and obey the same unitarity cuts, we do not expect exact agreement because of the freedom of generalized gauge invariance~\cite{GravYM2}. Regardless, we show that the discrepancy in the amplitudes vanish upon integration.
%%%%%%%%%%%%%%%%%%%%%%%%%%%%%%%%%%%%%%%%%%%%%%%
\subsection{$\mathcal{N}=4$ Super-Yang-Mills BCJ Numerators}
\label{sec:N4}
 %%%%%%%%%%%%%% FIGURE %%%%%%%%%%%%%
 \begin{figure}
 % Uses numbers instead of letters for subfloats
\renewcommand*\thesubfigure{\arabic{subfigure}}
    \subfloat[]{%
      \includegraphics[scale=.75]{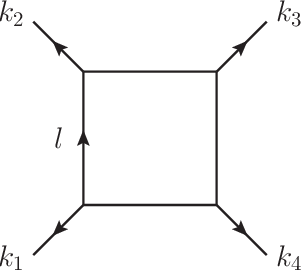}
    }
    \hspace{1cm}
    \subfloat[]{%
      \includegraphics[scale=.75]{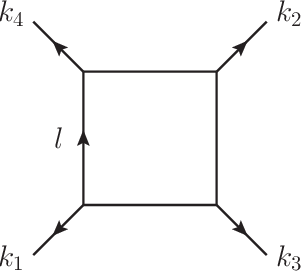}
    }
    \hspace{1cm}
    \subfloat[]{%
      \includegraphics[scale=.75]{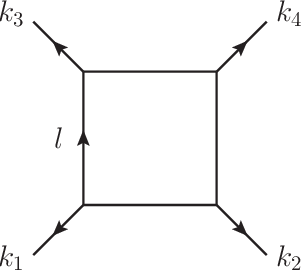}
    }
    \\[.5cm]
    \subfloat[]{%
      \includegraphics[scale=.75]{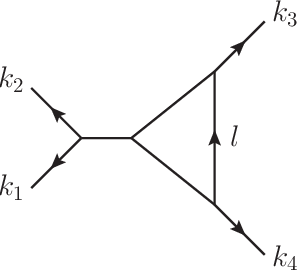}
    }
    \hspace{1cm}
    \subfloat[]{%
      \includegraphics[scale=.75]{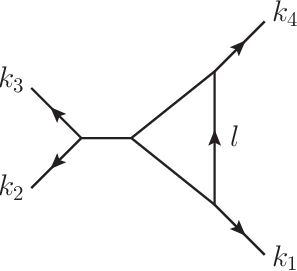}
    }
    \hspace{1cm}
    \subfloat[]{%
      \includegraphics[scale=.75]{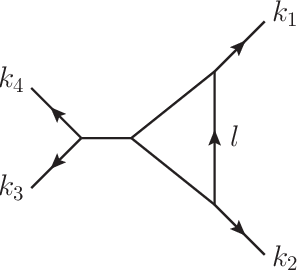}
    }
    \\[.5cm]
    \subfloat[]{%
      \includegraphics[scale=.75]{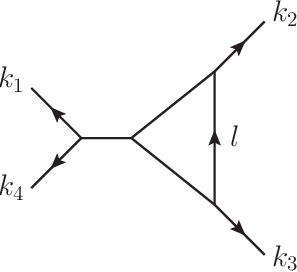}
    }
    \hspace{1cm}
    \subfloat[]{%
      \includegraphics[scale=.75]{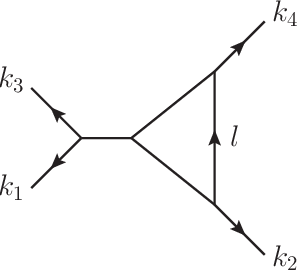}
    }
    \hspace{1cm}
    \subfloat[]{%
      \includegraphics[scale=.75]{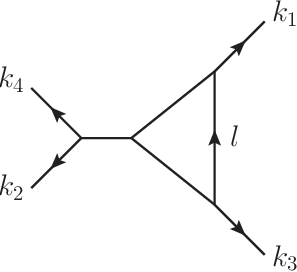}
    }
    \\[.75cm]
    \subfloat[]{%
      \includegraphics[scale=.75]{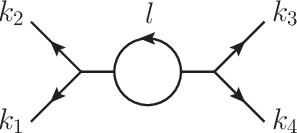}
    }
    \hspace{1cm}
    \subfloat[]{%
      \includegraphics[scale=.75]{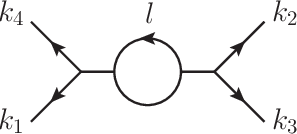}
    }
    \hspace{1cm}
    \subfloat[]{%
      \includegraphics[scale=.75]{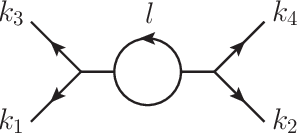}
    }
    \caption{The labeling convention for numerators with polarization vectors put into a momentum basis. The labeling is identical to that of Ref.~\cite{JJRadu}.}
    \label{fig:JJLabel}
  \end{figure}
 %%%%%%%%%%%%%%%%%%%%%%%%%%%%%%%%
As a warm-up exercise, we find duality-satisfying kinematic numerators in a helicity basis for $\mathcal{N}=4$ sYM theory. We do not immediately exploit the simplicity of the one-loop $\mathcal{N}=4$ box numerator, namely that it is proportional to the tree amplitude. Thus, the procedure that we outline here can be used in the more complicated cases described in the following subsections.

In general, we lose the relabeling properties mentioned in \sect{sec:formal} when we convert formal polarization vectors to a helicity basis. So, we first relabel \eqn{eq:n4} to get the three independent box numerators with external-leg orderings $(1,2,3,4)$, $(1,4,2,3)$, and $(1,3,4,2)$. (The box numerator with external-leg ordering $(1,4,3,2)$, for example, is the same as  the box numerator with external-leg ordering $(1,2,3,4)$ up to a relabeling of the loop momentum.) Then, we make the polarization-vector substitutions of Eqs.~(\ref{eq:rep}). Solving the numerator Jacobi relations, we generate all other independent numerators. (Alternatively, we could find all of the numerators with formal polarization vectors first, then apply Eqs.~(\ref{eq:rep}).) In this and the following subsections, we refer to \fig{fig:JJLabel} for numerator labeling conventions. We use $n_{i}$ to denote the kinematic numerator corresponding to \fig{fig:JJLabel}($i$). 

For the all-plus- and single-minus-helicity configurations,
\begin{align}
h_{1}=\pm, \hspace{.5cm} h_{2}=+, \hspace{.5cm} h_{3}=+, \hspace{.5cm} h_{4}=+,
\end{align}
our $\mathcal{N} = 4$ numerators vanish identically,
\begin{align}
n_{\text{1-12}}=0.
\end{align}
For the MHV configuration,
\begin{align}
h_{1}=-, \hspace{.5cm} h_{2}=-, \hspace{.5cm} h_{3}=+, \hspace{.5cm} h_{4}=+,
\end{align}
our numerators are
\begin{align}
n_{\text{1-3}}&=-i s^{2}e^{2i\left(2\phi_{12}-\phi_{14}-\phi_{23}\right)} \nonumber \\
&=stA_{4}^{\text{tree}}(1^{-}, 2^{-}, 3^{+}, 4^{+}), \\
n_{\text{4-12}}&=0,
\end{align}
where $A^{\text{tree}}_{4}$ is the color-ordered four-point  tree amplitude. These results are expected since \eqn{eq:n4} was already identified as $stA^{\text{tree}}_{4}$ in \app{sec:num4} even before external helicities were specified. It is well known that the tree amplitude is nonvanishing only for the MHV configuration. Also, the crossing symmetry of $stA^{\text{tree}}_{4}(1^{-},2^{-},3^{+},4^{+})$ assures us that all three box numerators should be identical. Because the box numerators are identical and there is no loop momentum present---as discussed in \sect{sec:formal}---that might need to be relabeled, the numerator Jacobi equations show that the non-box numerators vanish. All of these results agree with Ref.~\cite{N4}.

%%%%%%%%%%%%%%%%%%%%%%%%%%%%%%%%%%%%%%%%%%%%%%%
\subsection{$\mathcal{N}=1\text{ (chiral)}$ Super-Yang-Mills MHV BCJ Numerators}
\label{sec:N1}
 
Using the same procedure as \sect{sec:N4}, we now construct $\mathcal{N}=1$ (chiral) numerators in the MHV configuration, similar to those in Ref.~\cite{JJRadu}. We again use the labeling convention of \fig{fig:JJLabel}, which is identical to the convention of Ref.~\cite{JJRadu}. Also, we extract a factor of $stA_{4}^{\text{tree}}(1^{-},2^{-},3^{+},4^{+})$ from our numerators~\cite{stAtree, ck4l, JJRadu}. We define the quantity $N_{i}$ by
\begin{align}
n_{i}=stA_{4}^{\text{tree}}(1^{-},2^{-},3^{+},4^{+})N_{i}.
\end{align}
The resulting $N_{i}$'s are then,
\begin{subequations}
\begin{align}
N_{1}&= \tfrac{1}{2s^{2}}\left(4\tau_{51}\tau_{53} + 4\tau_{52}\tau_{53} - 2t \tau_{51} + 2u\tau_{52} + s\tilde{\tau}_{55}\right)+\tfrac{\mu^{2}}{s},
\vspace{.8cm}  \\[.25cm]
N_{2} &= \tfrac{1}{2s^{2}}\left(4\tau_{51}\tau_{53} + 4\tau_{52}\tau_{53} + 2s \tau_{53} + s\tilde{\tau}_{55}\right)+\tfrac{2i}{s^{2}}\left(l\cdot v\right)+\tfrac{\mu^{2}}{s},
\vspace{.8cm}  \\[.25cm]
N_{3}&= \tfrac{1}{2s^{2}}\left(4\tau_{51}\tau_{53} + 4\tau_{52}\tau_{53} + 2s \tau_{53} + s\tilde{\tau}_{55}\right)+\tfrac{\mu^{2}}{s},
\vspace{.8cm}  \\[.25cm]
N_{5}&= -\tfrac{2i}{s^{2}}\left(l\cdot v\right),
\vspace{.8cm}  \\[.25cm]
N_{7} &= \tfrac{1}{s^{2}}\left(t \tau_{51} - u\tau_{52} - s\tau_{53}\right) + \tfrac{2i}{s^{2}}\left(l\cdot v\right),
\vspace{.8cm}  \\[.25cm]
N_{8}&= \tfrac{2i}{s^{2}}\left(l\cdot v\right),
\vspace{.8cm}  \\[.25cm]
N_{9} &=  \tfrac{1}{s^{2}}\left(t \tau_{51} - u\tau_{52} + s\tau_{53}\right) + \tfrac{2i}{s^{2}}\left(l\cdot v\right),
\vspace{.8cm}  \\[.25cm]
N_{4} &= N_{6}= N_{10} = N_{11}=N_{12}=0,
\end{align}
\label{eq:Ns}
\end{subequations}
where $v$, $\mu^{2}$, and $\tau_{5i}$ are as defined in Eqs.~(\ref{eq:v}),~(\ref{eq:mu}),~and (\ref{eq:tau}), respectively. These numerators have the same simplicity as those of Ref.~\cite{JJRadu}---which we denote $\tilde{N}_{i}$---but they do not match exactly. (In this section and the next, we use a tilde to denote the results constructed from existing literature, Ref.~\cite{JJRadu} in this case.) Furthermore, the exact numerators of Ref.~\cite{JJRadu} cannot be obtained simply by choosing different reference momenta in \eqn{eq:rep}. The differences between our box numerators and theirs are
\begin{subequations}
\begin{align}
\Delta N_{1}&=\frac{i}{2s^{2}tu}\left[stu(\tilde{\tau}_{55}+2\mu^{2})-4s^{2}\tau_{51}\tau_{52}-4u^{2}\tau_{51}\tau_{53}-4t^{2}\tau_{52}\tau_{53}+2tu^{2}\tau_{51}-2t^{2}u\tau_{52}\right],
\\[.2cm]
\Delta N_{2}&=\Delta N_{1},
\\[.2cm]
\Delta N_{3}&=\frac{i}{2s^{2}tu}\left[stu(\tilde{\tau}_{55}+2\mu^{2})-4s^{2}\tau_{51}\tau_{52}-4u^{2}\tau_{51}\tau_{53}-4t^{2}\tau_{52}\tau_{53}+2s^{2}t\tau_{51}-2st^{2}\tau_{53}\right].
\end{align}
\end{subequations}
(We use $\Delta$ to denote our result minus the result from literature, eg. $\Delta N_{i}\equiv N_{i}-\tilde{N}_{i}$.)

To verify that our numerators of \eqn{eq:Ns} produce the same physical result as those in Ref.~\cite{JJRadu}, we show that the three independent color-ordered amplitudes $A^{(1)}_{4}(1,2,3,4)$, $A^{(1)}_{4}(1,4,2,3)$, and $A^{(1)}_{4}(1,3,4,2)$ of \eqn{eq:CO} give the same integrated result. The differences in the integrands $I_{4}(1,2,3,4)$, $I_{4}(1,4,2,3)$, and $I_{4}(1,3,4,2)$, are
\begin{subequations}
\begin{align}
\Delta I_{4}(1^{-},2^{-},3^{+},4^{+})&=\frac{i s t A_{4}^{\text{tree}}(1^{-},2^{-},3^{+},4^{+})}{2s(s\:t)\prod_{i=1}^{4}p^{2}_{i}}\left[s\: p_{1}^{2}\:p_{2}^{2} + t\: p_{3}^{2}\:p_{4}^{2} + u\: p_{4}^{2}\:p_{1}^{2} + 2\:s\:t\:\mu^{2}\right],
\\[.5cm]
\Delta I_{4}(1^{-},4^{+},2^{-},3^{+})&= \frac{i s t A_{4}^{\text{tree}}(1^{-},2^{-},3^{+},4^{+})}{2s(t\:u)\prod_{i=1}^{4}p^{2}_{i}}\left[u\: p_{1}^{2}\:p_{2}^{2} + s\: p_{2}^{2}\:p_{3}^{2} + t\: p_{3}^{2}\:p_{4}^{2} + 2\:t\:u\:\mu^{2}\right],
\\[.5cm]
\Delta I_{4}(1^{-},3^{+},4^{+},2^{-})&= \frac{i s t A_{4}^{\text{tree}}(1^{-},2^{-},3^{+},4^{+})}{2s(s\:u)\prod_{i=1}^{4}p^{2}_{i}}\left[u\: p_{2}^{2}\:p_{3}^{2} + t\: p_{3}^{2}\:p_{4}^{2} + s\: p_{4}^{2}\:p_{1}^{2} + 2\:s\:u\:\mu^{2}\right],
\end{align}
\end{subequations}
where we use the labeling convention of \fig{fig:ampLabel} so that two-particle cut-free terms may be readily identified (see \fig{fig:cuts}). We notice that, aside from the $\mu^{2}$ term, this difference does not survive either of the two-particle cuts. Integrals of this type---bubble-on-external-leg integrals sans the on-shell intermediate propagator, as in \eqn{eq:bubInt}---vanish in dimensional regularization (see Ref.~\cite{Smirnov}). Furthermore, the $\mu^{2}$ box integral does not contribute because it is $\mathcal{O}(\varepsilon)$~\cite{BernMorgan}. Hence, our integrated color-ordered amplitudes agree with those of Ref.~\cite{JJRadu}.
%%%%%%%%%%%%%%%%%%%%%%%%%%%%%%%%%%%%%%%%%%%%%%

%%%%%%%%%%%%%%%%%%%%%%%%%%%%%%%%%%%%%%%%%%%%%
\subsection{$\mathcal{N}=0$ Yang-Mills All-Plus-Helicity BCJ Numerators}

The contribution from a real scalar in the loop of a four-point one-loop amplitude has been computed in Ref.~\cite{BernMorgan}, where the external gluons are in the all-plus-helicity configuration. In the all-plus-helicity sector, the color-ordered amplitude for a gluon in the loop is simply twice that of a massless scalar,
\begin{align}
\label{eq:N0}
 \tilde{A}_{4}^{(1)\:\text{gluon}}\left(1^{+},2^{+},3^{+},4^{+}\right)=\dfrac{\left[ 12 \right]\left[ 34 \right]}{\left< 12 \right>\left< 34 \right>}\int\frac{d^{4}p}{(2\pi)^{4}}\frac{d^{-2\varepsilon}\mu}{(2\pi)^{-2\varepsilon}}\frac{2\: \mu^{4}}{\prod_{i=1}^{4}p^{2}_{i}},
\end{align}
where again we use a tilde to denote the results from existing literature.
Because the spinor-helicity prefactor and $\mu^{4}$ are invariant under relabelings of the external momenta and the loop momentum, we can immediately read off the BCJ numerators:
\begin{subequations}
\label{eq:ns0}
\begin{align}
\tilde{n}_{\text{1-3}}&=\frac{\left[ 12 \right]\left[ 34 \right]}{\left< 12 \right>\left< 34 \right>}\:2\:\mu^{4},
\\[.25cm]
\label{eq:ns0b}
\tilde{n}_{\text{4-12}}&=0,
\end{align}
\end{subequations}
where we again use the labeling conventions of \fig{fig:JJLabel}. (We identify $\frac{\left[ 12 \right]\left[ 34 \right]}{\left< 12 \right>\left< 34 \right>}2\mu^{4}$ as a box numerator because the four propagators present are those of the box diagram.)

Now, we construct all-plus-helicity $\mathcal{N}=0$ BCJ numerators in the same way as Sections~\ref{sec:N4} and~\ref{sec:N1}. However, we find that each box numerator in the all-plus-helicity sector is just as long as the formal-polarization expression; there is no simplification like we observed in \sect{sec:N1}. Furthermore, the non-box numerators do not vanish as in \eqn{eq:ns0b}. The color-ordered amplitude integrands, too, are more complicated than \eqn{eq:N0}. For instance,  
 \begin{align}
  \lefteqn{ \hskip 0 cm 
  \Delta A_{4}^{(1)\:\text{gluon}}\left(1^{+},2^{+},3^{+},4^{+}\right)=\dfrac{\left[ 12 \right]\left[ 34 \right]}{\left< 12 \right>\left< 34 \right>}\int\frac{d^{4}p}{(2\pi)^{4}}\frac{d^{-2\varepsilon}\mu}{(2\pi)^{-2\varepsilon}}\frac{1}{\left(\prod_{i=1}^{4}p^{2}_{i}\right) 6 s^{2}\:t^{2}\:(s+t)}\times}
  \vspace{.2cm} \nonumber \\ 
& \null\Bigl[
s \left(-3 s^2-2 t s+t^2\right) p_3^2 \:p_4^6
+s t (t-5 s) \:p_1^2 \:p_4^6
+\left(3 s^3+t s^2+t^2 s+3 t^3\right) p_3^4 \:p_4^4
 \vspace{.2cm} \nonumber \\ 
& \null
+s \left(6 s^2+11 t s+9 t^2\right) p_2^2 \:p_3^2 \:p_4^4
-2 t \left(-3 s^2+4 t s+3 t^2\right) p_1^2 \:p_3^2 \:p_4^4
+s t (9 s+17 t) \:p_1^2 \:p_2^2 \:p_4^4
 \vspace{.2cm} \nonumber \\ 
& \null
+3 t^2 (t-3 s) \:p_1^4 \:p_4^4
+t \left(s^2-2 t s-3 t^2\right) p_3^6 \:p_4^2
-2 s \left(3 s^2+4 t s+3 t^2\right) p_2^2 \:p_3^4 \:p_4^2
\vspace{.2cm} \nonumber \\ 
& \null
-s t (3 s+t) \:p_1^2 \:p_3^4 \:p_4^2
-s^2 (3 s+7 t) \:p_2^4 \:p_3^2 \:p_4^2
-2 s t (7 s+13 t) \:p_1^2 \:p_2^2 \:p_3^2 \:p_4^2
+t^2 (17 s+9 t) \:p_1^4 \:p_3^2 \:p_4^2
\vspace{.2cm} \nonumber \\ 
& \null
-2 s^2 t \:p_1^2 \:p_2^4 \:p_4^2
-8 s t^2 \:p_1^4 \:p_2^2 \:p_4^2
-6 t^3 \:p_1^6 \:p_4^2
+s t (s+t) \:p_2^2 \:p_3^6
+3 s^2 (s+t) \:p_2^4 \:p_3^4
\vspace{.2cm} \nonumber \\ 
& \null
+s t (5 s+9 t) \:p_1^2 \:p_2^2 \:p_3^4
+4 s^2 t \:p_1^2 \:p_2^4 \:p_3^2
+4 s t^2 \:p_1^4 \:p_2^2 \:p_3^2
-12 i \left(s^2-t^2\right) (l\cdot v) \:p_3^2 \:p_4^4
\vspace{.2cm} \nonumber \\ 
& \null
+12 i (s-t) t (l\cdot v) \:p_1^2 \:p_4^4
+12 i \left(s^2-t^2\right) (l\cdot v) \:p_3^4 \:p_4^2
+4 i s (3 s+7 t) (l\cdot v) \:p_2^2 \:p_3^2 \:p_4^2
\vspace{.2cm} \nonumber \\ 
& \null
-4 i t (5 s+3 t) (l\cdot v) \:p_1^2 \:p_3^2 \:p_4^2
+8 i s t (l\cdot v) \:p_1^2 \:p_2^2 \:p_4^2
+24 i t^2 (l\cdot v) \:p_1^4 \:p_4^2
-12 i s (s+t) (l\cdot v) \:p_2^2 \:p_3^4
\vspace{.2cm} \nonumber \\ 
& \null
-16 i s t (l\cdot v) \:p_1^2 \:p_2^2 \:p_3^2
+s t \left(7 s^2+6 t s-t^2\right) p_3^2 \:p_4^4
+s (11 s-t) t^2 \:p_1^2 \:p_4^4
\vspace{.2cm} \nonumber \\ 
& \null
+s t \left(-s^2+6 t s+7 t^2\right) p_3^4 \:p_4^2
-s^2 t (5 s+t) \:p_2^2 \:p_3^2 \:p_4^2
-s t^2 (13 s+5 t) \:p_1^2 \:p_3^2 \:p_4^2
+8 s^2 t^2 \:p_1^2 \:p_2^2 \:p_4^2
\vspace{.2cm} \nonumber \\ 
& \null
+12 s t^3 \:p_1^4 \:p_4^2
-s^2 t (s+t) \:p_2^2 \:p_3^4
-4 s^2 t^2 \:p_1^2 \:p_2^2 \:p_3^2
+24 i s t (s+t) (l\cdot v) \:p_3^2 \:p_4^2
\vspace{.2cm} \nonumber \\ 
& \null
-24 i s t^2 (l\cdot v) \:p_1^2 \:p_4^2
-4 s^2 t^2 (s+t) \:p_3^2 \:p_4^2
+6 s t^2 \left(2 s \mu ^2+2 t \mu ^2-s t\right) p_1^2 \:p_4^2
\Bigr],
\end{align}
where we again use the labeling convention of \fig{fig:ampLabel} so that two-particle cut-free terms may be readily identified. (We do not include contributions from bubble-on-external-leg diagrams since we demanded that they integrate to zero, as discussed in \sect{sec:formal}.) There are no terms that survive either two-particle cut of \fig{fig:cuts}. We then argue, as we did in \sect{sec:N1}, that this difference vanishes after integration. The tensor integrals involving $p\cdot v$ present no additional complications. By Lorentz invariance, the only objects that can contract with the dual vector, $v^{\mu}$, after integration are external momenta. These scalar products vanish. The other color-ordered amplitudes agree after integration, as well. Even though our amplitudes agree with literature, converting the formal-polarization numerators into a helicity basis using the method of \sect{sec:MomBasis} yields rather complicated terms that then integrate to zero.  Of course, these terms can be dropped immediately upon 
encountering them because they contain no $s$- or $t$-channel cuts.
%%%%%%%%%%%%%%%%%%%%%%%%%%%%%%%%%%%%%%%%%%

%%%%%%%%%%%%%%%%%%%%%%%%%%%%%%%%%%%%%%%%%%
\section{Conclusions}
\label{sec:Conclusions}
In Ref.~\cite{nonSUSYBCJ}, a representation for the one-loop four-point amplitude of pure Yang-Mills theory was constructed with the duality between color and kinematics manifest. In this paper, we extended the discussion by finding BCJ representations with fermions and scalars circulating in the loop. The presented expressions are valid in arbitrary dimensions and are written in terms of formal polarization vectors. Knowing the contributions from matter in the loop allowed us to construct supersymmetric BCJ amplitudes with external gluons. Furthermore, we found representations with improved loop-momentum power counting when supersymmetric field content is present.

We then compared a subset of our results to three amplitudes in
literature that obey color-kinematics duality: the $\mathcal{N}=4$ sYM
amplitude of Ref.~\cite{N4}, the $\mathcal{N}=1$ (chiral) MHV sYM
amplitude of Ref.~\cite{JJRadu}, and the $\mathcal{N}=0$
all-plus-helicity YM amplitude of Ref.~\cite{BernMorgan}. These
amplitudes were expressed in a four-dimensional helicity basis.  Our
formal-polarization BCJ numerators are lengthy in comparison to
previously obtained BCJ numerators given in four-dimensional helicity
representations, so it was interesting to see what simplifications
occur with helicity bases.

The $\mathcal{N}=4$ formal-polarization numerators
were identified as $stA_{4}^{\text{tree}}$, so we immediately obtained
the well-known results. Putting our $\mathcal{N}=1$ (chiral)
numerators into a four-dimensional MHV configuration revealed a
simplification on par with Ref.~\cite{JJRadu}. While these numerators
are not in exact agreement with Ref.~\cite{JJRadu}, we showed that
both sets of BCJ numerators produced the same amplitudes after
integration. Likewise, our $\mathcal{N}=0$ numerators in the
all-plus-helicity configuration produced the same integrated amplitude
as Ref.~\cite{BernMorgan}. However, our all-plus-helicity numerators
contained reasonably complicated terms that vanish on the unitarity
cuts.  For the generic case of nonsupersymmetric amplitudes there does
not appear to be any such simplification.

In summary, we provided further examples showing that BCJ duality
appears to extend to loop level even without supersymmetry.  It would
be important, not only to construct further loop-level examples, but to find a
systematic means of constructing loop integrands in a form compatible with
BCJ duality without relying on an ansatz.

%%%%%%%%%%%%%%%%%%%%%%%%%%%%%%%%%%%%%%%%%%%

%%%%%%%%%%%%%%%%%%%%%%%%%%%%%%%%%%%%%%%%%%%%
\subsection*{Acknowledgments}

I would like to express my gratitude to Z.~Bern and S.~Davies for their countless insights throughout the course of this work. I would also like to thank T.~Dennen, M.~S\o gaard, E.~Serna Campillo,  J.~J.~M.~Carrasco, and R.~Roiban for helpful discussions along the way. Finally, I would like acknowledge J.~Stankowicz, S.~Litsey, and A.~Sivaramakrishnan for their input in the preparation of this paper. This material is based upon work supported by the Department of Energy under Award Number \mbox{DE--SC0009937}.
%%%%%%%%%%%%%%%%%%%%%%%%%%%%%%%%%%%%%%%%%%%%

\appendix
%%%%%%%%%%%%%%%%%%%%%%%%%%%%%%%%%%%%%%%%%%%
\section{SUSY BCJ Box Numerators}
\label{sec:num4}
In this appendix, we provide the four-dimensional $\mathcal{N}=4$ and $\mathcal{N}=1$ (chiral) BCJ box numerators with formal polarization vectors. As discussed in \sect{sec:formal}, supersymmetry reduces the maximum power of loop momentum in our box numerators. Specifically, it is reduced from $\mathcal{O}(p^{4})$ in $\mathcal{N}=0$  to $\mathcal{O}(p^{2})$ in $\mathcal{N}=1$ and $\mathcal{O}(p^{0})$ in $\mathcal{N}=4$. While clearly seen in the $\mathcal{N}=4$ expression below, this property is not explicit in the $\mathcal{N}=1$ (chiral) expression. The improved power counting is only manifest when the inverse propagators in the numerators are expanded. For example, observing the labeling convention of \fig{fig:ampLabel} for the box numerator with external-leg ordering $(1,2,3,4)$, we would have to expand $p_{3}^{2}$ as so,
\begin{align}
p^{2}_{3} = (p-k_{1}-k_{2})^{2} = p^{2}-2\:(p\cdot k_{1})-2\:(p \cdot k_{2}) + s^{2}.
\end{align}

The $\mathcal{N}=4$ BCJ box numerator in four dimensions is as follows:
%%%%% N=4 %%%%%%
\begin{align}
\label{eq:n4}
 \lefteqn{ \hskip 0 cm 
n^{\mathcal{N}=4}_{1234;p}\: =  -i \Bigl[
-\tfrac{1}{4}\:\mathcal{E}_{13}\:\mathcal{E}_{24}\:s\:t
+\tfrac{1}{2}\:\mathcal{E}_{14}\:\mathcal{E}_{23}\:s\:t
+\tfrac{1}{2}\:\mathcal{E}_{14}\:\mathcal{E}_{23}\:s^2
-\mathcal{E}_{13}\:\mathcal{K}_{24}\:\mathcal{K}_{42}\:s}
\vspace{.2cm} \nonumber \\ 
& \null
-\mathcal{E}_{24}\:\mathcal{K}_{23}\:\mathcal{K}_{41}\:s
-2\:\mathcal{E}_{12}\:\mathcal{K}_{23}\:\mathcal{K}_{34}\:s
-\mathcal{E}_{13}\:\mathcal{K}_{12}\:\mathcal{K}_{34}\:s
-2\:\mathcal{E}_{23}\:\mathcal{K}_{24}\:\mathcal{K}_{31}\:s
\vspace{.2cm} \nonumber \\ 
& \null
-2\:\mathcal{E}_{12}\:\mathcal{K}_{23}\:\mathcal{K}_{24}\:s
-2\:\mathcal{E}_{13}\:\mathcal{K}_{12}\:\mathcal{K}_{24}\:s
-2\:\mathcal{E}_{14}\:\mathcal{K}_{12}\:\mathcal{K}_{23}\:s
-2\:\mathcal{E}_{14}\:\mathcal{K}_{12}\:\mathcal{K}_{13}\:s
\Bigr]
+\mathrm{cyclic}.
\end{align}
%%%%%%%%%%%%%%
Even with formal polarization vectors, we can identify this as $stA^{\text{tree}}_{4}(1,2,3,4)$.

The $\mathcal{N}=1$ (chiral) BCJ box numerator in four dimensions is 
%%%%%%N=1%%%%%%
\begin{align}
\label{eq:n1}
 \lefteqn{ \hskip 0 cm 
n^{\mathcal{N}=1\text{(chiral)}}_{1234;p}\: =  -i \Bigl[
-\tfrac{1}{4}\:\mathcal{E}_{12}\:\mathcal{E}_{34}\:p^{2}_1\:p^{2}_3
+\tfrac{1}{4}\:\mathcal{E}_{13}\:\mathcal{E}_{24}\:p^{2}_1\:p^{2}_3
+\tfrac{1}{4}\:\mathcal{E}_{12}\:\mathcal{E}_{34}\:p^{2}_1\:p^{2}_2
-\tfrac{1}{4}\:\mathcal{E}_{13}\:\mathcal{E}_{24}\:p^{2}_1\:p^{2}_2}
\vspace{.2cm} \nonumber \\ 
& \null
+\tfrac{1}{4}\:\mathcal{E}_{14}\:\mathcal{E}_{23}\:p^{2}_1\:p^{2}_2
-\tfrac{1}{4}\:\mathcal{E}_{14}\:\mathcal{E}_{23}\left(p^{2}_1\right)^2
-\mathcal{E}_{23}\:\mathcal{K}_{41}\:\mathcal{P}_{44}\:p^{2}_1
+\mathcal{E}_{12}\:\mathcal{K}_{34}\:\mathcal{P}_{33}\:p^{2}_1
-\mathcal{E}_{24}\:\mathcal{K}_{31}\:\mathcal{P}_{33}\:p^{2}_1
\vspace{.2cm} \nonumber \\ 
& \null
+\mathcal{E}_{34}\:\mathcal{K}_{41}\:\mathcal{P}_{22}\:p^{2}_1
+\mathcal{E}_{34}\:\mathcal{K}_{31}\:\mathcal{P}_{22}\:p^{2}_1
+\mathcal{E}_{13}\:\mathcal{K}_{24}\:\mathcal{P}_{22}\:p^{2}_1
-\mathcal{E}_{23}\:\mathcal{K}_{34}\:\mathcal{P}_{11}\:p^{2}_1
-\mathcal{E}_{23}\:\mathcal{K}_{24}\:\mathcal{P}_{11}\:p^{2}_1
\vspace{.2cm} \nonumber \\ 
& \null
-\tfrac{1}{4}\:\mathcal{E}_{12}\:\mathcal{E}_{34}\:p^{2}_2\:s
+\tfrac{1}{4}\:\mathcal{E}_{13}\:\mathcal{E}_{24}\:p^{2}_2\:s
-\tfrac{1}{4}\:\mathcal{E}_{14}\:\mathcal{E}_{23}\:p^{2}_2\:s
+\tfrac{1}{4}\:\mathcal{E}_{14}\:\mathcal{E}_{23}\:p^{2}_1\:s
-\tfrac{1}{2}\:\mathcal{E}_{13}\:\mathcal{P}_{22}\:\mathcal{P}_{44}\:s
\vspace{.2cm} \nonumber \\ 
& \null
-\tfrac{1}{2}\:\mathcal{E}_{24}\:\mathcal{P}_{11}\:\mathcal{P}_{33}\:s
+\mathcal{E}_{34}\:\mathcal{P}_{11}\:\mathcal{P}_{22}\:s
+\tfrac{1}{2}\:\mathcal{E}_{34}\:\mathcal{K}_{41}\:\mathcal{K}_{42}\:p^{2}_1
+\tfrac{1}{2}\:\mathcal{E}_{34}\:\mathcal{K}_{31}\:\mathcal{K}_{42}\:p^{2}_1
+\tfrac{1}{2}\:\mathcal{E}_{13}\:\mathcal{K}_{24}\:\mathcal{K}_{42}\:p^{2}_1
\vspace{.2cm} \nonumber \\ 
& \null
-\tfrac{1}{2}\:\mathcal{E}_{14}\:\mathcal{K}_{23}\:\mathcal{K}_{42}\:p^{2}_1
-\tfrac{3}{2}\:\mathcal{E}_{23}\:\mathcal{K}_{34}\:\mathcal{K}_{41}\:p^{2}_1
-\mathcal{E}_{23}\:\mathcal{K}_{24}\:\mathcal{K}_{41}\:p^{2}_1
+\tfrac{1}{2}\:\mathcal{E}_{34}\:\mathcal{K}_{12}\:\mathcal{K}_{41}\:p^{2}_1
-\tfrac{1}{2}\:\mathcal{E}_{23}\:\mathcal{K}_{31}\:\mathcal{K}_{34}\:p^{2}_1
\vspace{.2cm} \nonumber \\ 
& \null
+\tfrac{1}{2}\:\mathcal{E}_{12}\:\mathcal{K}_{23}\:\mathcal{K}_{34}\:p^{2}_1
-\tfrac{1}{2}\:\mathcal{E}_{23}\:\mathcal{K}_{24}\:\mathcal{K}_{31}\:p^{2}_1
-\tfrac{1}{2}\:\mathcal{E}_{24}\:\mathcal{K}_{23}\:\mathcal{K}_{31}\:p^{2}_1
+\tfrac{1}{2}\:\mathcal{E}_{34}\:\mathcal{K}_{12}\:\mathcal{K}_{31}\:p^{2}_1
\vspace{.2cm} \nonumber \\ 
& \null
+\tfrac{1}{2}\:\mathcal{E}_{13}\:\mathcal{K}_{12}\:\mathcal{K}_{24}\:p^{2}_1
-\tfrac{1}{2}\:\mathcal{E}_{14}\:\mathcal{K}_{12}\:\mathcal{K}_{23}\:p^{2}_1
-\mathcal{K}_{24}\:\mathcal{K}_{42}\:\mathcal{P}_{11}\:\mathcal{P}_{33}
+2\:\mathcal{K}_{23}\:\mathcal{K}_{34}\:\mathcal{P}_{11}\:\mathcal{P}_{22}
\vspace{.2cm} \nonumber \\ 
& \null
+2\:\mathcal{K}_{13}\:\mathcal{K}_{34}\:\mathcal{P}_{11}\:\mathcal{P}_{22}
+\tfrac{1}{2}\:\mathcal{E}_{34}\:\mathcal{K}_{41}\:\mathcal{P}_{22}\:s
-\tfrac{1}{2}\:\mathcal{E}_{13}\:\mathcal{K}_{34}\:\mathcal{P}_{22}\:s
-\tfrac{1}{2}\:\mathcal{E}_{13}\:\mathcal{K}_{24}\:\mathcal{P}_{22}\:s
+\tfrac{1}{2}\:\mathcal{E}_{14}\:\mathcal{K}_{23}\:\mathcal{P}_{22}\:s
\vspace{.2cm} \nonumber \\ 
& \null
+\tfrac{1}{2}\:\mathcal{E}_{34}\:\mathcal{K}_{42}\:\mathcal{P}_{11}\:s
+\tfrac{1}{2}\:\mathcal{E}_{23}\:\mathcal{K}_{34}\:\mathcal{P}_{11}\:s
+\tfrac{1}{2}\:\mathcal{E}_{23}\:\mathcal{K}_{24}\:\mathcal{P}_{11}\:s
-\tfrac{1}{2}\:\mathcal{E}_{24}\:\mathcal{K}_{23}\:\mathcal{P}_{11}\:s
+\tfrac{1}{2}\:\mathcal{E}_{34}\:\mathcal{K}_{12}\:\mathcal{P}_{11}\:s
\vspace{.2cm} \nonumber \\ 
& \null
+\mathcal{K}_{23}\:\mathcal{K}_{34}\:\mathcal{K}_{42}\:\mathcal{P}_{11}
-\mathcal{K}_{23}\:\mathcal{K}_{24}\:\mathcal{K}_{42}\:\mathcal{P}_{11}
-\mathcal{K}_{13}\:\mathcal{K}_{24}\:\mathcal{K}_{42}\:\mathcal{P}_{11}
-\mathcal{K}_{12}\:\mathcal{K}_{23}\:\mathcal{K}_{24}\:\mathcal{P}_{11}
\vspace{.2cm} \nonumber \\ 
& \null
-\mathcal{K}_{12}\:\mathcal{K}_{13}\:\mathcal{K}_{24}\:\mathcal{P}_{11}
-\tfrac{1}{16}\:\mathcal{E}_{13}\:\mathcal{E}_{24}\:s\:t
+\tfrac{1}{8}\:\mathcal{E}_{14}\:\mathcal{E}_{23}\:s\:t
-\tfrac{1}{4}\:\mathcal{E}_{13}\:\mathcal{K}_{24}\:\mathcal{K}_{42}\:s
+\tfrac{1}{2}\:\mathcal{E}_{23}\:\mathcal{K}_{24}\:\mathcal{K}_{41}\:s
\vspace{.2cm} \nonumber \\ 
& \null
-\tfrac{1}{4}\:\mathcal{E}_{24}\:\mathcal{K}_{23}\:\mathcal{K}_{41}\:s
-\tfrac{1}{4}\:\mathcal{E}_{13}\:\mathcal{K}_{12}\:\mathcal{K}_{34}\:s
-\tfrac{1}{2}\:\mathcal{E}_{13}\:\mathcal{K}_{12}\:\mathcal{K}_{24}\:s
+\tfrac{1}{2}\:\mathcal{E}_{14}\:\mathcal{K}_{12}\:\mathcal{K}_{23}\:s
-\tfrac{1}{4}\:\mathcal{K}_{13}\:\mathcal{K}_{24}\:\mathcal{K}_{31}\:\mathcal{K}_{42}
\vspace{.2cm} \nonumber \\ 
& \null
-\tfrac{1}{2}\:\mathcal{K}_{12}\:\mathcal{K}_{23}\:\mathcal{K}_{34}\:\mathcal{K}_{41}
-\mathcal{K}_{12}\:\mathcal{K}_{23}\:\mathcal{K}_{31}\:\mathcal{K}_{34}
-\tfrac{1}{2}\:\mathcal{K}_{12}\:\mathcal{K}_{13}\:\mathcal{K}_{31}\:\mathcal{K}_{34}
-\mathcal{K}_{12}\:\mathcal{K}_{23}\:\mathcal{K}_{24}\:\mathcal{K}_{31}
\vspace{.2cm} \nonumber \\ 
& \null
-\mathcal{K}_{12}\:\mathcal{K}_{13}\:\mathcal{K}_{24}\:\mathcal{K}_{31}
\Bigr]
+\mathrm{cyclic}.
\end{align}
%%%%%%%%%%%%%%%%%%%%%%%%%%%%%%%%%%%%%%%%%%%

%%%%%%%%%%%%%%%%%%%%%%%%%%%%%%%%%%%%%%%%%%%
\section{Determination of the Phase Factor}
\label{sec:phase}
We determine the phase factor, $\mathcal{P}_{h_{i}}(i;j_{1},j_{2})$, through comparison with the spinor-helicity representation of the polarization vectors (cf. Ref.~\cite{Dixon}):
\begin{align}
\varepsilon_{\pm}^{\mu}(i; j_{1})\equiv\pm\frac{\left\langle j_{1}^{\mp}\right|\gamma^{\mu}\left| i^{\mp}\right\rangle}{\sqrt{2}\left\langle j_{1}^{\mp} | i^{\pm}\right\rangle}.
\end{align}
Using the spin sum completeness relation in the massless limit,
\begin{align}
\slashed{p} = \sum_{s=1,2} u_{s}(p)\bar{u}_{s}(p)=u_{+}(p)\bar{u}_{+}(p)+u_{-}(p)\bar{u}_{-}(p)=\left| p^{+} \right\rangle \left\langle p^{+} \right| + \left| p^{-} \right\rangle\left\langle p^{-} \right|,
\end{align}
and
\begin{align}
\left\langle i^{-} | j^{+} \right\rangle \equiv \left\langle i j \right\rangle, \hspace{1cm}
\left\langle i^{+} | j^{-} \right\rangle \equiv \left[ i j \right],  \hspace{1cm}
\left\langle i^{+} | j^{+} \right\rangle=0, \hspace{1cm}
\left\langle i^{-} | j^{-} \right\rangle=0,
\end{align}
we find, for $k_{j_{2}}\neq k_{i} \neq k_{j_{1}}$,
\begin{align}
k_{j_{2}}\cdot\varepsilon_{h_{i}}(i; j_{1})
=\pm\frac{\left\langle j_{1}^{\mp}\right|\slashed{k}_{j_{2}}\left| i^{\mp}\right\rangle}{\sqrt{2}\left\langle j_{1}^{\mp} | i^{\pm}\right\rangle}
=\begin{cases} +\frac{\left\langle j_{1} j_{2} \right\rangle \left[ j_{2} i \right]}{\sqrt{2}\left\langle j_{1} i \right\rangle}, & \mbox{if } h_{i}\mbox{ is }+ \vspace{.2cm}\\ -\frac{\left[ j_{1} j_{2} \right] \left\langle j_{2} i \right\rangle}{\sqrt{2}\left[ j_{1} i \right]}, & \mbox{if } h_{i}\mbox{ is }- \end{cases}
=-\sqrt{\frac{s_{j_{1}j_{2}}s_{j_{2}i}}{2\: s_{j_{1} i}}}e^{-i h_{i} \left(\phi_{ij_{2}}-\phi_{j_{1}j_{2}}+\phi_{i j_{1}}\right)}.
\label{eq:dot1}
\end{align}
Without loss of generality, we compare this result to the inner product of $k_{j_{2}}$ with the momentum basis representation of \eqn{eq:epDef},
\begin{align}
k_{j_{2}}\cdot\varepsilon_{h_{i}}(i;j_{1},j_{2})
=\mathcal{P}_{h_{i}}(i;j_{1},j_{2})\sqrt{\frac{s_{j_{1}j_{2}}s_{j_{2}i}}{2 s_{j_{1} i}}},
\label{eq:dot2}
\end{align}
to conclude that
\begin{align}
\mathcal{P}_{h_{i}}(i;j_{1},j_{2})=-e^{-i h_{i} \left(\phi_{ij_{2}}-\phi_{j_{1}j_{2}}+\phi_{i j_{1}}\right)}.
\end{align}
Since there are only three independent external momenta, there is nothing special about choosing $k_{j_{2}}$ as both the second reference momentum in \eqn{eq:dot2} and as the momentum to contract with the polarization vectors in Eqs.~(\ref{eq:dot1}) and~(\ref{eq:dot2}). The other choice leads to the same result: there is a minus sign from momentum conservation on the external legs and also a minus sign due to momentum conservation in the phase factors from Eqs.~(\ref{eq:phi}).
%%%%%%%%%%%%%%%%%%%%%%%%%%%%%%%%%%%%%%%%%

%%%%%%%%%%%%%%%%%%%%%%%%%%%%%%%%%%%%%%%%%%%%%%

\end{document}